\documentclass[twocolumn,pra,amsmath,amssymb,superscriptaddress,longbibliography,nofootinbib,floatfix]{revtex4-2}
\usepackage[utf8]{inputenc}
\usepackage{pgfplots}
\usepackage{pgfplotstable}
\usepackage{bbold}
\usepackage{comment}
\usepackage[colorlinks=true, linkcolor=red, allbordercolors={white}]{hyperref}
\usepackage{float}
\usepackage{subfiles}
\usepackage{cleveref}
\pgfplotsset{compat = newest}
\usepgfplotslibrary{colorbrewer}
\usepgfplotslibrary{groupplots}
\usetikzlibrary{arrows.meta}
\definecolor{color1}{HTML}{1f77b4}
\definecolor{color2}{HTML}{ff7f0e}
\definecolor{color3}{HTML}{2ca02c}
\definecolor{color4}{HTML}{d62728}
\definecolor{color5}{HTML}{9467bd}
\definecolor{color6}{HTML}{8c564b}
\definecolor{color7}{HTML}{e377c2}
\definecolor{color8}{HTML}{7f7f7f}
\definecolor{color9}{HTML}{bcbd22}
\definecolor{shadyblue}{rgb}{0.33,0.33,1}
\colorlet{fillLambda}{gray!30}
\colorlet{fillGamma}{gray!60}
\pgfplotsset{
  cycle list={color1\\color2\\color3\\color4\\color5\\color6\\color7\\color8\\color9\\},
}
\newcommand{\overbar}[1]{\mkern 1.5mu\overline{\mkern-1.5mu#1\mkern-1.5mu}\mkern 1.5mu}

\newcommand{\arrowIn}{
\tikz \draw[-{Stealth[length=2mm, width=1.5mm]}] (-1pt,0) -- (1pt,0);
}
\tikzset{
    new dash/.code args={on #1 off #2}{
        \csname tikz@addoption\endcsname{%
            \pgfgetpath\currentpath%
            \pgfprocessround{\currentpath}{\currentpath}%
            \csname pgf@decorate@parsesoftpath\endcsname{\currentpath}{\currentpath}%
            \pgfmathparse{\csname pgf@decorate@totalpathlength\endcsname-#1}\let\rest=\pgfmathresult%
            \pgfmathparse{#1+#2}\let\onoff=\pgfmathresult%
            \pgfmathparse{max(floor(\rest/\onoff), 1)}\let\nfullonoff=\pgfmathresult%
            \pgfmathparse{max((\rest-\onoff*\nfullonoff)/\nfullonoff+#2, #2)}\let\offexpand=\pgfmathresult%
            \pgfsetdash{{#1}{\offexpand}}{0pt}}%
    }
}

\usepackage{letltxmacro}
\LetLtxMacro{\originaleqref}{\eqref}
\renewcommand{\eqref}{Eq.~\originaleqref}

\crefname{equation}{Eq.}{Eqs.}

\providecommand{\norm}[1]{\lVert#1\rVert}

\begin{document}

\title{Characterizing the Entanglement of Anyonic Systems\\using the Anyonic Partial Transpose}

\author{Nico Kirchner}
%\email{nico.kirchner@tum.de}
\affiliation{Technical University of Munich, TUM School of Natural Sciences, Physics Department, 85748 Garching, Germany}
\affiliation{Munich Center for Quantum Science and Technology (MCQST), Schellingstr. 4, 80799 M{\"u}nchen, Germany}
\author{Wonjune Choi}
\affiliation{Technical University of Munich, TUM School of Natural Sciences, Physics Department, 85748 Garching, Germany}
\affiliation{Munich Center for Quantum Science and Technology (MCQST), Schellingstr. 4, 80799 M{\"u}nchen, Germany}
\author{Frank Pollmann}
\affiliation{Technical University of Munich, TUM School of Natural Sciences, Physics Department, 85748 Garching, Germany}
\affiliation{Munich Center for Quantum Science and Technology (MCQST), Schellingstr. 4, 80799 M{\"u}nchen, Germany}

\begin{abstract}
Entanglement of mixed quantum states can be quantified using the partial transpose and its corresponding entanglement measure, the logarithmic negativity. Recently, the notion of partial transpose has been extended to systems of anyons, which are exotic quasiparticles whose exchange statistics go beyond the bosonic and fermionic case. Studying the fundamental properties of this anyonic partial transpose, we first reveal that when applied to the special case of fermionic systems, it can be reduced to the fermionic partial transpose or its twisted variant depending on whether or not a boundary Majorana fermion is present. Focusing on ground state properties, we find that the anyonic partial transpose captures both the correct entanglement scaling for gapless systems, as predicted by conformal field theory, and the phase transition between a topologically trivial and a nontrivial phase. For non-abelian anyons and the bipartition geometry, we find a rich multiplet structure in the eigenvalues of the partial transpose, the so-called negativity spectrum, and reveal the possibility of defining both a charge- and an imbalance-resolved negativity.
\end{abstract}

\maketitle

\section{Introduction}

While elementary particles in our three-dimensional universe are either bosons or fermions, two-dimensional quantum matter permits more exotic varieties of emergent quasiparticles called anyons~\cite{Leinaas_identical, Wilczek_anyon, Wilczek_anyon2, anyon_review}.
These anyons are point-like excitations of two-dimensional intrinsic topological order (TO) and possess nontrivial braiding statistics, a manifestation of the long-ranged entanglement, which serves as a defining feature setting TO apart from conventional phases that lack such intricate order.
To understand TO in realistic solid-state systems, such as fractional quantum Hall states~\cite{Laughlin_FQH, Halperin_FQH, Arovas_FQH, Wen_FQH, FQH_review} and potential topological spin liquids~\cite{Wen_TO, Kitaev_TC, Kitaev_anyon, Savary_QSLreview}, accounting for the effects of finite temperature is essential~\cite{finiteTnegativity_toric_code, Grover_topological_negativity_finiteT, Grover_entanglement_length, Grover_singularity_negativity, Liu_QHfiniteTLN}.
Furthermore, fueled by the success of synthetic quantum matter in cold atom platforms, there is a strong interest in non-thermal mixed states~\cite{Bao_mixedTO, Fan_mixedTO, Lee_decoherenceSPT, FloquetTO}.
Since prepared quantum states deviate from equilibrium and contend with various sources of decoherence, investigating mixed-state entanglement measures for TO is indispensable.

While the standard measure for diagnosing bipartite entanglement in a pure state is the entanglement entropy \cite{area_law_EE}, its applicability diminishes when extending the analysis to more general mixed states because it accounts for both quantum entanglement and classical fluctuations within the statistical mixture.
In contrast, the logarithmic entanglement negativity (LN), which quantifies the violation of the positive partial transpose criterion \cite{BPT_Peres, BPT_Horodecki}, is more suitable for mixed-state entanglement since the LN selectively measures quantum correlations encoded in the density operator while disregarding classical uncertainty \cite{BPT_Vidal, LN_Plenio, Eisler_2015, FPT_Shapourian_Ryu, FPT2_Shapourian_Ryu, APT_Shapourian_Ryu, Calabrese_QFTnegativity, Calabrese_finiteTnegativity, Shapourian_finiteTnegativity}.

Although the LN itself has been explored for TO \cite{Lee_LN_TO, LN_TC, LN_CS_bulk, LN_CS_edge, Liu_QHfiniteTLN, Grover_topological_negativity_finiteT, CCNR_TO, RVB_separability, APT_Shapourian_Ryu, APT_TO_Ryu, Bao_mixedTO, Fan_mixedTO}, most previous studies have concentrated on the ground state and low-lying excited states of TO, typically involving only a few numbers of anyons.
However, in this context, there is yet another important class of systems: novel quantum phases of matter emerging from the interaction of many anyons \cite{GoldenChain2, Fibonacci_Collective, Spin1Models2, GoldenChain, Spin1Models, PhysRevB.90.075129, anyonic_defects}.
When anyons are in close proximity, the residual interactions among them can give rise to the nucleation of new phases, such as gapped topological phases \cite{PhysRevB.90.075129, Spin1Models, Spin1Models2} or gapless phases described by conformal field theory (CFT) \cite{GoldenChain, Fibonacci_Collective, Spin1Models, Spin1Models2, anyonic_defects}, superimposed upon the original parent TO.
In particular, a one-dimensional chain of interacting anyons is constrained by non-invertible topological symmetries rooted in the non-trivial fusion rules of non-Abelian anyons \cite{GoldenChain, Spin1Models, anyonic_defects}.
These unique symmetries can prevent relevant perturbations that could otherwise gap out the spectrum in fermionic or bosonic systems.
Thus, rich families of (1+1)D CFTs can be stabilized in interacting anyon chains, offering a plethora of emergent phases. These phases can be considered bona fide states within the bulk or representative exotic gapless boundary modes of topological phases. We substantiate our theoretical insights through numerical studies in conjunction with our broader discussion on interacting anyons and their entanglement structure.
Specifically, we focus on a one-dimensional array of anyons with nearest-neighbor interactions, providing a concrete and illustrative context for validating and showcasing our findings.

% Summary
The primary objective of this paper is to unravel the inherent structure of the generalized definition of partial transpose for anyonic systems, introduced in Ref.~\cite{APT_Shapourian_Ryu}, and to explore its implications, e.g., on the complex-valued eigenvalue spectrum of the partially transposed anyonic density operator, called the negativity spectrum \cite{negativity_Hamiltonian, NegSpectrumCFT, NegSpectrumGapped, NegSpectrumRandomMixed, NegSpectra_TwistedAndUntwisted}.
A crucial subtlety in understanding LN arises from its diverse definitions for states involving many particles \cite{BPT_Vidal, LN_Plenio, Eisler_2015, FPT_Shapourian_Ryu, APT_Shapourian_Ryu}.
Due to the nontrivial exchange statistics of anyons, the partial transpose introduces nontrivial phase factors and intricate mixing of anyonic charges.
Therefore, a sophisticated redefinition of the partial transpose is essential, contingent upon the chosen \emph{local} degrees of freedom constituting quantum phases.
Furthermore, non-Abelian anyons possess nontrivial fusion rules, rendering the Fock space of many-anyon systems more complex than the straightforward tensor product structure seen in bosonic or fermionic systems.
Considering various fusion channels becomes imperative even for noninteracting sets of many-anyon states, unveiling a nontrivial structure absent in bosonic or fermionic models.
Here, we elucidate various aspects of the anyonic partial transpose, investigate the LN of gapped and critical phases in a one-dimensional chain of interacting anyons, propose a generalization of the imbalance-resolved negativity \cite{ChargeResolvedNegativity, ChargeResolvedNegativity2} for many-anyon systems, and reveal the possibility of defining a charge-resolved negativity for bipartitions.
The exploration promises to unveil the rich and distinctive features inherent in the entanglement structure of anyonic systems, transcending the conventional paradigms of bosons and fermions.

% Organization
The paper is structured as follows.
In Section~\ref{sec:PartialTranspose}, we provide a concise overview of the partial transpose and the LN for bosons and fermions, elucidating its generalization to a system of anyons as introduced in Ref.~\cite{APT_Shapourian_Ryu}.
Section~\ref{sec:RelationAPT-FPT} establishes a concrete connection between the fermionic partial transpose, its twisted version and the anyonic partial transpose when applied to fermionic systems.
In Section~\ref{sec:LowEnergy}, we explore how LN can characterize the low-energy properties of a critical state described by CFT and the topological phase transition between two gapped phases in one dimension.
For a bipartite system of anyons, Sec.~\ref{sec:NegativitySpectrum} discusses the generic block diagonal structure of the anyonic partial transpose of the density operator $\tilde{\rho}^{T_A^a}$ and explores how the braiding statistics correlates with the complex eigenvalue spectrum of $\tilde{\rho}^{T_A^a}$.
To validate our findings, numerical studies are conducted on the golden chain and the $S=1$ anyonic $\mathrm{su}(2)_5$ spin chain.
Motivated by the block diagonal structure observed in $\tilde{\rho}^{T_A^a}$, Sec.~\ref{sec:ChargeImbalanceResolvedALN} introduces the definition of charge- and imbalance-resolved anyonic negativity.
The conclusion and a roadmap for future research are presented in Sec.~\ref{sec:Conclusion}.
Technical details concerning numerical calculations and some analytical results are provided in the appendices.

\section{Partial Transposes}
\label{sec:PartialTranspose}

Let us start with a quick review of the notion of separability and motivate why the different notions of partial transpose are necessary. We then briefly introduce the bosonic partial transpose (BPT)~\cite{BPT_Peres, BPT_Horodecki, BPT_Vidal} and the fermionic partial transpose (FPT)~\cite{FPT_Shapourian_Ryu, FPT2_Shapourian_Ryu, FermionReflectionOperator, FPT3_Shapourian_Ryu, choi2023, Eisler_2015} and finally discuss the anyonic partial transpose (APT)~\cite{APT_Shapourian_Ryu, APT_TO_Ryu}.

The different notions of partial transpose are suited to analyze the entanglement in mixed states between two generic subsystems. Determining the entanglement then generally involves partial traces over all subsystems one is not interested in such that the partial transpose can be applied to the mixed state associated with the two subsystems of interest. Throughout this work, we focus on the bipartition geometry, which means in the context of the definitions of the partial transposes below and Sec.~\ref{sec:RelationAPT-FPT} that the corresponding partial traces have already been applied. In all other contexts, we assume that the geometry of the \emph{full system} is bipartite such that no partial traces are involved. This constraint is purely geometrical in the sense that the state describing the system may still be mixed.

%The different notions of partial transpose are suited to analyze the entanglement in mixed states between two generic subsystems. Such general geometries then require tracing out of all other subsystems, such that the partial transpose can be applied to the mixed state associated with the two subsystems of interest. Throughout this work, we focus on the bipartition geometry, which in the case of the definitions of the partial transposes below and Sec.~\ref{sec:RelationAPT-FPT} means that the corresponding partial traces have already been performed. In all other cases however, this means that the geometry of the \emph{full system} is bipartite, i.e., no partial traces have been performed. The state describing the system may nevertheless be mixed.

\subsubsection*{Separability}

In order to discuss the mixed-state entanglement between two parties, we first need to provide a precise definition.
Two subsystems, $A$ and $B$, are considered to be separable (or not entangled) if a density operator $\rho$ describing the quantum state of $A \cup B$ can be expressed as a convex combination of tensor products of local density operators $\rho_A^{(i)}$ and $\rho_B^{(i)}$ describing subsystems $A$ and $B$:
\begin{align}
\rho = \sum_i p_i \rho_A^{(i)} \otimes \rho_B^{(i)},
\end{align}
where $p_i > 0$ and $\sum_i p_i = 1$.
This definition is motivated by the fact that a separable state can be prepared using only local operations and classical communications (LOCCs).
If a mixed state is not separable, the two subsystems are considered to be entangled, and the state requires additional quantum resources to be prepared.

For bosonic systems, the local density operators $\rho_A^{(i)}$ and $\rho_B^{(i)}$ are required to be positive semidefinite Hermitian operators with unit trace.
However, in fermionic or anyonic systems, which exhibit nontrivial many-body exchange statistics, the definition of separability requires additional qualifications.

For fermionic systems, the \emph{physical} density operators $\rho_A^{(i)}$ and $\rho_B^{(i)}$, which can be prepared via LOCCs, must conserve fermion number parity~\cite{FPT_Shapourian_Ryu, FPT2_Shapourian_Ryu}.
This implies that the local density operators cannot represent a quantum superposition of states with an even and odd number of fermions.
Similarly, in anyonic systems, a superselection rule needs to be imposed, requiring that the \emph{physical}, locally preparable density operators must conserve the subsystems' total topological charges.
Thus, the local quantum state can be mixed but cannot be a superposition of states with different topological charges.

Therefore, even if the total system's density operator $\rho$ can be formally decomposed into a convex combination of positive semidefinite Hermitian operators $\rho_A^{(i)}$ and $\rho_B^{(i)}$, the quantum state may not be separable if the Hermitian operators $\rho_A^{(i)}$ and $\rho_B^{(i)}$ do not preserve fermion number parity for fermionic systems~\cite{FPT_Shapourian_Ryu, FPT2_Shapourian_Ryu} or the topological charge for anyonic systems.
This additional constraint necessitates a generalized definition of the partial transpose to quantify the mixed-state entanglement in fermionic and, more generally, anyonic many-body systems.

\subsubsection*{Bosonic Partial Transpose}

\begin{figure}[t]
\centering
\includegraphics{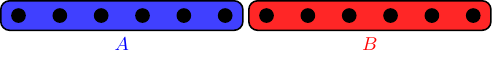}
\caption{Bipartition of a chain into two subsystems $A$ and $B$.}
\label{fig:bipartition}
\end{figure}

Consider a bipartition of the physical system (after optional partial traces) into two subsystems $A$ and $B$ as depicted in Fig.~\ref{fig:bipartition}. Then, density matrices can be expressed as
\begin{align}
	\rho = \sum_{ijkl} \rho_{ijkl} \left|\phi_A^{(i)},\phi_B^{(j)}\right\rangle \left\langle \phi_A^{(k)},\phi_B^{(l)} \right|,
\end{align}
where $\rho_{ijkl}=\langle \phi_A^{(i)},\phi_B^{(j)}| \rho | \phi_A^{(k)},\phi_B^{(l)} \rangle$ are the density matrix elements and $\lbrace |\phi_{A,B}^{(i)}\rangle\rbrace$ denote orthonormal bases of the respective subsystems. The BPT $\rho^{T_A^b}$ of $\rho$ is obtained by exchanging the states in $A$~\cite{BPT_Peres, BPT_Horodecki, BPT_Vidal}
\begin{align}
	\rho^{T_A^b} = \sum_{ijkl} \rho_{ijkl} \left|\phi_A^{(k)},\phi_B^{(j)}\right\rangle \left\langle \phi_A^{(i)},\phi_B^{(l)} \right|.
	\label{eq:BPT}
\end{align}
In order to quantify the entanglement of $\rho$, we can define the logarithmic entanglement negativity (LN) $\mathcal{E}$~\cite{LN_Plenio, BPT_Vidal},
\begin{align}
	\mathcal{E}_b = \ln (\norm{\rho^{T_A^b}}_1),
	\label{eq:BEN}
\end{align}
where $\norm{A}_1= \mathrm{Tr}\sqrt{AA^\dagger}=\mathrm{Tr}\sqrt{A^\dagger A}$ denotes the trace norm. The LN vanishes for separable states, which thus represents a necessary (not a sufficient) condition for separability. As the name suggests, the BPT can be applied to bosonic systems, such as spin chains for example.%\footnote{In principle, \eqref{eq:BPT} can also be applied to fermionic systems~\cite{Eisler_2015}, there are however some problems associated with doing so. First of all, there are some unfavorable properties such as $(\rho^{T_A})^{T_B} \neq \rho^{T_{A\cup B}}$ and $(\rho_1 \otimes \ldots \otimes \rho_n)^{T_A} \neq \rho_1^{T_A} \otimes \ldots \otimes \rho_n^{T_A} $~\cite{FPT_Shapourian_Ryu, FPT2_Shapourian_Ryu, FPT3_Shapourian_Ryu}. More importantly, it has been shown that entangled fermionic states with zero bosonic LN and nonzero fermionic LN exist~\cite{FPT2_Shapourian_Ryu}, implying that only the latter correctly captures the entanglement. Thus, the FPT introduced in \eqref{eq:FPT} should be used for fermionic systems.}.

%\footnote{In principle, \eqref{eq:BPT} can be applied to fermionic systems~\cite{Eisler_2015}, the disadvantage of doing so are some unfavorable properties such as $(\rho^{T_A})^{T_B} \neq \rho^{T_{A\cup B}}$ and $(\rho_1 \otimes \ldots \otimes \rho_n)^{T_A} \neq \rho_1^{T_A} \otimes \ldots \otimes \rho_n^{T_A} $~\cite{FPT_Shapourian_Ryu, FPT2_Shapourian_Ryu, FPT3_Shapourian_Ryu}, which can be avoided by using the FPT introduced in \eqref{eq:FPT}.}.

\subsubsection*{Fermionic Partial Transpose}

For fermionic systems, nontrivial phase factors arise when performing the partial transpose. Consider the occupation basis $|\lbrace n_j\rbrace_A,\lbrace n_j\rbrace_B\rangle$, where $\lbrace n_j\rbrace_{A,B}$ denotes the occupation within the respective subsystem. The FPT $\rho^{T_A^f}$ of $\rho$ can be defined in this basis via~\cite{FPT_Shapourian_Ryu, FPT2_Shapourian_Ryu, FPT3_Shapourian_Ryu}
\begin{align}
	\begin{split}
	&\big(\left|\lbrace n_j\rbrace_A,\lbrace n_j\rbrace_B \right\rangle  \left\langle\lbrace \overbar{n}_j\rbrace_A, \lbrace\overbar{n}_j\rbrace_B\right|\big)^{T_A^f} \\
	&= (-1)^{\phi(\lbrace n_j\rbrace,\lbrace \overbar{n}_j\rbrace)} \left|\lbrace \overbar{n}_j\rbrace_A,\lbrace n_j\rbrace_B \right\rangle  \left\langle\lbrace n_j\rbrace_A, \lbrace\overbar{n}_j\rbrace_B\right|,
	\end{split}
	\label{eq:FPT}
\end{align}
with the nontrivial phase factor
\begin{align}
	\begin{split}
	\phi(\lbrace n_j\rbrace,\lbrace \overbar{n}_j\rbrace) = \frac{\tau_A(\tau_A+2)}{2}+\frac{\overbar{\tau}_A(\overbar{\tau}_A+2)}{2}+\tau_B\overbar{\tau}_B \\
	+ \tau_A\tau_B + \overbar{\tau}_A\overbar{\tau}_B + (\tau_A + \tau_B)(\overbar{\tau}_A + \overbar{\tau}_B).
	\end{split}
	\label{eq:FermionPhaseFactor}
\end{align}
Here, $\tau_X=\sum_{j\in X}n_j$ and $\overbar{\tau}_X=\sum_{j\in X}\overbar{n}_j$ denote the occupation numbers in subsystem $X$, with $X\in \lbrace A,B\rbrace$. Analogous to bosonic systems, the FPT can be used to compute the fermionic LN, which is given by $\mathcal{E}_f=\ln (\norm{\rho^{T_A^f}}_1)$.

\begin{figure}[b]
	\includegraphics{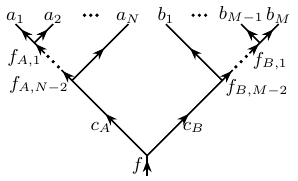}
	\caption{Basis choice for describing bipartitions of the system into two subsystems $A$ and $B$ containing $N$ and $M$ anyons, respectively. The anyonic charges are denoted by $a_i$, $b_i$, their fusion products by $f_{A,i}$, $f_{B,i}$, with $c_A\equiv f_{A,N-1}$, $c_B\equiv f_{B,M-1}$, and the total topological charge of the system by $f$.}
	\label{fig:AnyonBasis}
\end{figure}

\subsubsection*{Anyonic Partial Transpose}

Let us now focus on the partial transpose for anyonic systems, for which the formalism of fusion diagrams is used. For an introduction to this topic see, e.g., Refs.~\cite{APT_Shapourian_Ryu, simon2020topological, 1506.05805, 0707.4206, bonderson_2007, 2102.05677, InsideOutsideBases, PhysRevB.107.195129}. For a good minimal introduction containing all necessary details for the APT, see Ref.~\cite{APT_Shapourian_Ryu}. We assume from here on that the reader is acquainted with the basic notions of these diagrammatics. Note that in this work, we focus on the case $N_{ab}^c\in \lbrace 0,1\rbrace$ for simplicity, i.e., we do not consider higher fusion multiplicities.

For concreteness, let us choose the diagrams depicted in Fig.~\ref{fig:AnyonBasis} as a particular basis from the very beginning to describe bipartitions of the system (cf. Fig.~\ref{fig:bipartition}), where $a_i(b_i)$ denotes the anyons in subsystem $A(B)$ containing $N(M)$ anyons, $f_{A,i}(f_{B,i})$ the respective fusion products and $f$ the total charge of the system. The basis was chosen in a way such that the total fusion product of each subsystem can be directly read off; this makes the computation of the APT comparably easy, as shown below. For convenience, let us denote the fusion product of subsystem $A(B)$ as $c_A(c_B)\equiv f_{A,N-1}(f_{B,M-1})$. For a density matrix\footnote{We refer to density matrices of anyonic systems as $\widetilde{\rho}$. The reason for this is that there are two different ways to normalize anyonic density matrices, which are usually distinguished by refering to them as $\rho$ and $\widetilde{\rho}$. See Refs.~\cite{APT_Shapourian_Ryu, APT_TO_Ryu, 0707.4206, bonderson_2007, InsideOutsideBases} for details.} $\widetilde{\rho}$ describing some anyonic system, the APT $\widetilde{\rho}^{T_A^a}$ can be defined in the basis in Fig.~\ref{fig:AnyonBasis} as~\cite{APT_Shapourian_Ryu, APT_TO_Ryu}

\begin{widetext}
\begin{align}
	\begin{split}
	\widetilde{\rho}^{T_A^a}&=\sum_{\mathbf{x}, \mathbf{x'},f}  \frac{p\big(\mathbf{x},\mathbf{x'}; f\big)}{N_f} \left(\hspace{-12.5pt}\vcenter{\includegraphics{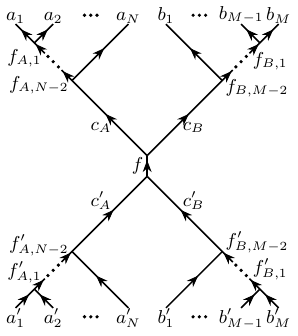}}\hspace{-360pt}\right)^{T_A^a}
	\hspace{-10.5pt}=\sum_{\mathbf{x}, \mathbf{x'},f}  \frac{p\big(\mathbf{x}, \mathbf{x'};f\big)}{N_f} \hspace{-5pt} \vcenter{\includegraphics{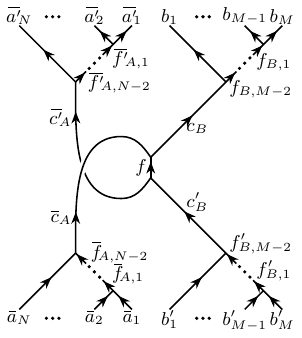}}\hspace{-380pt}
	\end{split}\label{eq:APT_def}
\end{align}
\begin{minipage}{.5\textwidth}
\begin{align}
	\begin{split}
	&=\sum_{\mathbf{x},\mathbf{x'},g}  \frac{d_g M\big(\mathbf{x},\mathbf{x'}; g\big)}{N_g} \hspace{-15pt}\vcenter{\includegraphics{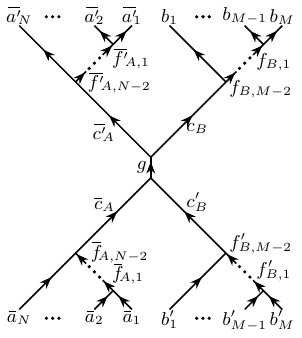}}\hspace{-150pt}
	\end{split}\label{eq:APT_sol}
\end{align}
\end{minipage}%
\begin{minipage}{.5\textwidth}
\begin{align}
	&\text{with}\quad N_f = \bigg[\Big(\prod_{i\in A} d_{a_i}d_{a_i'}\Big)\Big(\prod_{j\in B} d_{b_j}d_{b_j'}\Big)d_f^2\bigg]^{\frac{1}{4}},\label{eq:APT_normalization}\\
	\begin{split}
	&M(\mathbf{x},\mathbf{x'}; g) = \sum_f \frac{p\big(\mathbf{x},\mathbf{x'}; f\big)}{\sqrt{d_{c_B}d_{c_B'}}} A^{c_Ac_B}_f \Big(A^{c_A'c_B'}_f\Big)^*\\
	&\qquad \qquad \qquad \qquad\times R^{\overline{c}_Ac_B'}_g \Big(R^{f\overline{c}_A}_{c_B}\Big)^* \Big[F_g^{\overbar{c'}_{\hspace{-2pt}A}f\overline{c}_A}\Big]_{c_B'c_B},\hspace{-50pt}
	\end{split}
	\label{eq:APT_sol2}
\end{align}
\end{minipage}
\end{widetext}
where $d_e$ denotes the quantum dimension of charge $e$ and $\overline{e}$ its anticharge. We further introduced the superindices $\mathbf{x}$ and $\mathbf{x'}$ that contain all charges of the ket and bra parts of each contribution to $\widetilde{\rho}$. That is, $\mathbf{x}$ is a superindex for $\lbrace a_i \rbrace$, $\lbrace b_i \rbrace$, $\lbrace f_{A,i} \rbrace$ and $\lbrace f_{B,i} \rbrace$; $\mathbf{x'}$ similarly contains on all primed anyons. The normalization condition of the density matrix, $\widetilde{\mathrm{Tr}}\widetilde{\rho} = 1$, implies that
\begin{align}
	\sum_{\mathbf{x},f} p\big(\mathbf{x},\mathbf{x}; f\big) = 1.
\end{align}
Note the similarity of the result in \eqref{eq:APT_sol} to the dimer case discussed in Ref.~\cite{APT_Shapourian_Ryu}, which we rederive in App.~\ref{app:APTDimer} for a self-contained illustration. The fusion products of the subsystems ($c_A,c_A',c_B,c_B'$) take the role of the corresponding dimer charges. In addition to that, the anyons in subsystem $A$ are exchanged between the bra and ket and their configurations are spatially reversed compared to $\widetilde{\rho}$. The APT as defined above is not unique since we may choose to braid the anyons either clockwise or counterclockwise. This choice does not affect the anyonic LN (ALN), which is defined analogously to the bosonic and fermionic LN as $\mathcal{E}_a=\ln (\norm{\widetilde{\rho}^{T_A^a}}_1)$. Thus, this choice is irrelevant for determining the entanglement. 

%The above notions of the partial transpose in \cref{eq:BPT,eq:FPT,eq:APT_def} can of course also be applied to non-bipartite systems by first partially tracing until two subsystems remain and then applying the respective definitions.

%\section{Properties of the Anyonic Partial Transpose}

%Having introduced the APT, let us now discuss some interesting properties it possesses. We start by relating the APT applied to fermionic systems to the FPT.

\section{Relation between the Anyonic and the Fermionic Partial Transpose}
\label{sec:RelationAPT-FPT}

The aim of this section is to relate the APT applied to fermionic systems to the FPT. As discussed below, we find two relations, one connecting the APT to the FPT and one relating the APT to the \emph{twisted} FPT~\cite{NegSpectra_TwistedAndUntwisted}. The essential difference between these two relations lies in the chosen boundary conditions, that is, the presence or absence of a boundary Majorana fermion. These results suggest that the APT is a proper generalization of the partial transpose to anyonic systems since it can reproduce the limiting cases.

Before stating the relations, we note that there is some intuition where the correspondence between the APT and the FPT comes from. In Ref.~\cite{APT_Shapourian_Ryu}, it was pointed out that the FPT of a Majorana fermion dimer density matrix is very similar to the exchange operator of two vortices in $p_x+ip_y$ superconductors, suggesting that the nontrivial phases in the FPT can be interpreted as braid operations of underlying Majoranas. In the definition of the APT in \eqref{eq:APT_def}, we can see such braid operations between the particles in subsystem $A$ (the braid operation involving the total charges can be translated to braid operations among the individual particles, see the definition of the APT in Ref.~\cite{APT_Shapourian_Ryu}). This intuitively suggests that the APT applied to fermionic systems and the FPT may be related.

%In Ref.~\cite{APT_Shapourian_Ryu}, it was pointed out that the FPT of a Majorana fermion dimer density matrix is very similar to the exchange operator of two vortices in $p_x+ip_y$ superconductors, suggesting that the nontrivial phases of the FPT can be interpreted as braid operations of underlying Majoranas. While this point of view provides some intuition as to where the phases in the FPT may come from, the question whether and how precisely the FPT and the APT applied to fermionic systems are related is still unanswered. This is, however, a very important question: On the one hand, the definition of the APT in \eqref{eq:APT_def} suggests that particles are exchanged, just like in the interpretation above. On the other hand, one would expect the anyonic version to reproduce the fermionic case if it is a true generalization of the partial transpose.

\subsubsection*{Relation to the Twisted Fermionic Partial Transpose}
The APT can be related to the FPT via
\begin{align}
	\widetilde{\rho}^{T^a_A} = \mathcal{R}_A\rho^{T^f_A}(-1)^{F_A}\mathcal{R}_A^{-1},
	\label{eq:ConnectionAPT-FPT}
\end{align}
where $(-1)^{F_A}$ is the fermionic parity operator in subsystem $A$, i.e., $F_A = \sum_{j\in A}n_j$, and $\mathcal{R}_A$ is the spatial reflection operator restricted to $A$, with $\mathcal{R} f_j \mathcal{R}^{-1} = if_{L+1-j}$~\cite{FermionReflectionOperator} (the lattice is labeled by $j=1,\ldots,L$). Note that $\rho^{T^f_A}(-1)^{F_A}$ has already been studied and is called the \emph{twisted} FPT~\cite{NegSpectra_TwistedAndUntwisted}. The derivation of this relation can be found in App.~\ref{app:FPTandAPT}.

While this result already suggests that the APT is a generalization of both the BPT and the FPT to anyonic systems when it comes to the LN as entanglement measure, it has to be noted that the \emph{negativity spectrum}, i.e., the eigenvalues of $\widetilde{\rho}^{T^a_A}$ and $\rho^{T^f_A}$, do not agree with each other. However, the negativity spectrum of the APT $\widetilde{\rho}^{T^a_A}$ does agree with the one of the twisted FPT $\rho^{T^f_A}(-1)^{F_A}$ since the spatial reflection $\mathcal{R}_A$ in \eqref{eq:ConnectionAPT-FPT} is unitary.

\subsubsection*{Relation to the Fermionic Partial Transpose}
%The connection in \eqref{eq:ConnectionAPT-FPT} between the APT and twisted FPT does not rely on Majorana fermions at all, despite the intuition given in the first part of this section. This raises the question whether there is yet another connection that makes use of Majorana fermions. We can indeed find such a connection by imposing certain boundary conditions for the APT: Let $\widetilde{\rho}_{\sigma}$ denote the anyonic density matrix obtained from $\widetilde{\rho}$ (which only contains fermions as charges $a_i^{(\prime)}$ and $b_i^{(\prime)}$) by adding a Majorana fermion $\sigma$ as boundary condition to subsystem $A$, that is, we add an anyon of charge $\sigma$ as $a_0$ to the basis in \eqref{eq:AnyonBasis}. We can relate the APT of $\widetilde{\rho}_{\sigma}$ to the FPT:\footnote{For this relation, we ignore the boundary conditions for convenience, i.e., we identify the state containing a Majorana fermion as boundary charge in the APT with the state containing a trivial boundary charge if the two fermion configurations agree.}

The connection in \eqref{eq:ConnectionAPT-FPT} between the APT and twisted FPT does not rely on Majorana fermions at all, despite the intuition given in the first part of this section. It turns out that there is yet another connection that does make use of Majorana fermions by using them to impose certain boundary conditions for the APT: Let $\widetilde{\rho}_{\sigma}$ denote the anyonic density matrix obtained from $\widetilde{\rho}$ (which only contains fermions as charges $a_i^{(\prime)}$ and $b_i^{(\prime)}$) by adding a Majorana fermion $\sigma$ as boundary condition to subsystem $A$, that is, we add an anyon of charge $\sigma$ as $a_0$ to the basis in Fig.~\ref{fig:AnyonBasis}. We can relate the APT of $\widetilde{\rho}_{\sigma}$ to the FPT:\footnote{For this relation, we ignore the boundary conditions for convenience, i.e., we identify the state containing a Majorana fermion as boundary charge in the APT with the state containing a trivial boundary charge if the two fermion configurations agree.}
\begin{align}
	\widetilde{\rho}_{\sigma}^{T^a_A} = \frac{\theta_{\sigma}}{d_{\sigma}} U \mathcal{R}_A\rho^{T^f_A}\mathcal{R}_A^{-1}U^{\dagger}.
	\label{eq:ConnectionAPT-FPT2}
\end{align}
Here, we utilized the Ising anyon model~\cite{APT_Shapourian_Ryu, bonderson_2007} to describe the behavior of the Majorana fermion $\sigma$, see App.~\ref{app:FPTandAPT} for details. The factor $\theta_{\sigma}=e^{i\pi/8}$ denotes the topological twist of $\sigma$, which arises from the fact that $\widetilde{\mathrm{Tr}}(\widetilde{\rho}_{\sigma}^{T^a_A})=\theta_{\sigma}$; the renormalization by the quantum dimension $d_{\sigma}=\sqrt{2}$ can be traced back to the difference in normalization between the fermionic and the anyonic density matrix. Finally, the unitary transformation $U$ is given in the fermion occupation basis by
\begin{align}
	U \left|\lbrace n_j\rbrace_A,\lbrace n_j\rbrace_B \right\rangle = (-i)^{\tau_A \mathrm{mod}2} \left|\lbrace n_j\rbrace_A,\lbrace n_j\rbrace_B \right\rangle,
\end{align}
where we may interpret the factor $(-i)^{\tau_A \mathrm{mod}2}$ as coming from an exchange between the boundary Majorana fermion $\sigma$ and the total fusion product of subsystem $A$ without the boundary. An even $\tau_A$ corresponds to a trivial exchange with a boson and an odd $\tau_A$ corresponds to an exchange with a fermion $\psi$ with phase $R^{\sigma\psi}_{\sigma}=-i$. This interpretation is reminiscent of the spatial reflection $\mathcal{R}_A$ that is applied to subsystem $A$ and entails exchanges between the fermionic creation and annihilation operators. That is, we may interpret $U$ as adding the exchange phase of the boundary Majorana fermion with the rest of subsystem $A$, as one would expect for a spatial reflection. The derivation of this relation can be found in App.~\ref{app:FPTandAPT}.

With this second connection between the APT and the FPT, we have shown again that the APT is indeed the correct generalization of the PT to anyonic systems and that it can even reproduce the full spectrum of the FPT. It also follows that using different boundary charges may lead to different notions of the APT which may have different properties, such as the FPT and its twisted counterpart.

Note that in the case of \eqref{eq:ConnectionAPT-FPT2}, the total charge of the system corresponds to a Majorana fermion $\sigma$, which is only physical if there is another Majorana fermion in a subsystem that has already been traced over. This immediately raises the question whether it is possible to find yet another connection between the APT and the FPT when using Majorana fermions as boundary charges for both subsystem $A$ and $B$ such that the total charge is trivial again. Under this assumption, it is in fact \emph{not} possible to find such a connection. The easiest way to see this is by noting that the eigenvalues of the APT possess some multiplet structure, which will be explained in detail later in Sec.~\ref{sec:NegativitySpectrum}. This structure implies that when $\sigma$ is the boundary charge in both $A$ and $B$, there need to be pairwise degeneracies in the absolute values of the spectrum of the APT, which can be found neither in the usual nor in the twisted FPT.

%The derivation of this relation can be found in App.~\ref{app:FPTandAPT}. This result suggests that the APT is indeed a generalization of both the BPT and the FPT to anyonic systems when it comes to the LN as entanglement measure. However, it has to be noted that the \emph{negativity spectrum}, related to the eigenvalues of $\widetilde{\rho}^{T^a_A}$ and $\rho^{T^f_A}$, do not agree with each other. However, the negativity spectrum of the APT $\widetilde{\rho}^{T^a_A}$ does agree with the one of the twisted FPT $\rho^{T^f_A}(-1)^{F_A}$ since the spatial reflection $\mathcal{R}_A$ in \eqref{eq:ConnectionAPT-FPT} is unitary.

\section{Logarithmic Negativity of Critical and Topological Anyonic States}
\label{sec:LowEnergy}

Let us now focus on the low-energy properties of the ALN using the numerical approach of matrix product states (MPS) applied to anyonic systems~\cite{AnyonMPS, AnyonMPS2, AnyonMPS3, AnyonMPS4}. Details on the computation of the ALN using MPS are discussed in App.~\ref{app:MPS}.

\subsection{Anyonic Hamiltonians}

Throughout the remainder of this work, we will mainly focus on two models with bipartite geometry and open boundary conditions (OBC). The first model is known as golden chain~\cite{GoldenChain, GoldenChain2} and describes a chain of Fibonacci anyons fulfilling the fusion rules $1\times \tau=\tau \times 1=\tau$ and $\tau\times\tau = 1+\tau$, where $1$ denotes the trivial charge and $\tau$ the Fibonacci anyonic charge. The golden chain Hamiltonian favors the fusion of neighboring anyons to the identity channel,
\begin{align}
	\mathcal{H}_{\mathrm{GC}} = -\sum_{i=1}^{L-1} P_{i,i+1}^{(1)},
	\label{eq:GoldenChain}
\end{align}
where $P_{i,i+1}^{(1)}$ denotes the projector of the anyons located at sites $i$ and $i+1$ onto their trivial fusion channel. Interactions of this form arise from virtual tunneling processes of topological charges between the anyons, which corresponds to the leading order interactions of the effective theory for large distances between the anyons~\cite{Tunneling}. As seen from \eqref{eq:GoldenChain}, we use OBC and fix the boundary topological charges (additional boundary degrees of freedom corresponding to extending the basis in Fig.~\ref{fig:AnyonBasis} by $a_0$ and $b_{M+1}$, without $\mathcal{H}_{\mathrm{GC}}$ directly acting on them) to be trivial.

The second, slightly more complicated model features $\mathrm{su}(2)_k$ anyons~\cite{APT_Shapourian_Ryu, bonderson_2007, Spin1Models2, Spin1Models} with fusion rules
\begin{align}
	j_1\times j_2 = \sum_{j=|j_1-j_2|}^{\min \lbrace j_1+j_2,k-j_1-j_2 \rbrace} j,
\end{align}
where the charges take half-integer values up to $k/2$, i.e., $j_1,j_2 \in \lbrace 0,1/2,1,\ldots, k/2 \rbrace$. Note that in the context of $\mathrm{su}(2)_k$, we denote the trivial charge by $0$, in all other contexts we denote it by $1$. Using these anyons, we can build analogues to spin-$1$ systems subject to Hamiltonians whose local terms project onto the total spin sectors of two neighboring spins with certain amplitudes; one paradigmatic example of such a Hamiltonian is the AKLT Hamiltonian~\cite{HaldaneModel, AKLT, AKLT2}. The anyonic generalization to $\mathrm{su}(2)_k$ anyons features a chain of spin-$1$ anyons where the Hamiltonian consists of projectors $P^{(1)}_{i,i+1}$ and $P^{(2)}_{i,i+1}$ projecting the anyons on sites $i$ and $i+1$ onto their direct fusion products $1$ and $2$. The Hamiltonian can thus be written as~\cite{Spin1Models2, Spin1Models}
\begin{align}
	\mathcal{H}_{\mathrm{spin-}1}(\theta)=\sum_{i=1}^{L-1} \cos(\theta) P^{(2)}_{i,i+1} -\sin(\theta) P^{(1)}_{i,i+1},
	\label{eq:Spin1Model}
\end{align}
where we use again OBC and choose trivial boundary charges. While this Hamiltonian can be studied for each level $k$, we will focus on the case $k=5$.

\subsection{Logarithmic Entanglement Negativity Scaling for the Golden Chain Model}
\label{sec:NegativityScaling}

%In Sec.~\ref{sec:ChargeResolvedALN}, we derived a formula for the charge-resolved ALN and applied it to the ground state of a gapped system. Let us now extend the study of such low energy / ground state properties to gapless systems by considering the scaling of the ALN for the equal bipartition geometry over different system sizes $L$. For gapless system described by conformal field theory (CFT), the LN is expected to behave as~\cite{}
%\begin{align}
%	\mathcal{E} = \frac{c}{4}\ln(L) + \mathrm{const.},
%\end{align}
%where $c$ denotes the central charge of the CFT.

We first consider gapless systems described by CFTs, for which the ground states are known to show nontrivial entanglement behavior. In particular, when monitoring the growth of the entanglement entropy for finite systems over different system sizes or different bipartitions, it is possible to extract the central charge of the CFT~\cite{Vidal_QCPentanglement, Calabrese_EE_QFT, Calabrese_EE_CFT}. This is also possible using the LN, which is expected to follow~\cite{Calabrese_2013}
\begin{align}
	\mathcal{E} = \frac{c}{4}\ln\left(\frac{2L}{\pi}\right) + \mathrm{const.}
	\label{eq:NegativityScaling}
\end{align}
in the case of bosons, an equal bipartition and OBC, where $c$ denotes the central charge of the CFT and $L$ the system size. It has been further shown that \eqref{eq:NegativityScaling} is also expected to hold for fermionic systems~\cite{NegSpectra_TwistedAndUntwisted}, which raises the question whether it also holds true for anyonic systems. We provide numerical evidence in four different cases that this is indeed the case.

\begin{figure}[t]
	\centering
	\includegraphics{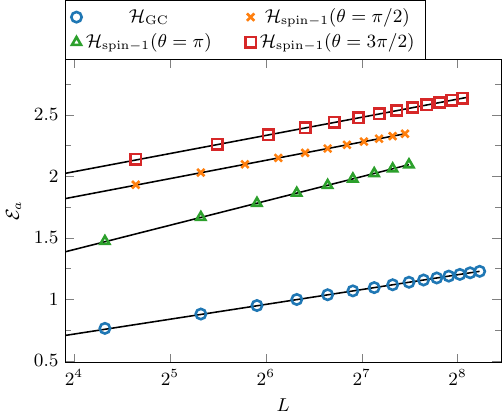}
	\caption{ALN of the ground states of the critical models $\mathcal{H}_{\mathrm{GC}}$, $\mathcal{H}_{\mathrm{spin-}1}(\theta=\pi /2)$, $\mathcal{H}_{\mathrm{spin-}1}(\theta=\pi)$ and $\mathcal{H}_{\mathrm{spin-}1}(\theta=3\pi /2)$ for OBC, equal bipartitions and different total system sizes $L$ as obtained from MPS simulations. Scaling analyses yield for the central charges $c=0.696(2)$, $c=0.8582(8)$, $c=1.1405(8)$ and $c=0.8507(6)$ for MPS bond dimension $\chi=500$, which are close to the known values of the CFTs $c=7/10$, $c=6/7$, $c=8/7$ and $c=6/7$, respectively. The results for $\mathcal{H}_{\mathrm{spin-}1}(\theta=3\pi /2)$ have been shifted for clarity.}
\label{fig:NegativityScaling}
\end{figure}

We use the golden chain $\mathcal{H}_{\mathrm{GC}}$ introduced in \eqref{eq:GoldenChain} and the anyonic spin-$1$ model for $\mathrm{su}(2)_5$ anyons $\mathcal{H}_{\mathrm{spin-}1}(\theta)$ introduced in \eqref{eq:Spin1Model} as test objects for extracting the correct central charges using the ALN. The golden chain model is known to be described by the tricritical Ising CFT with central charge $c=7/10$~\cite{GoldenChain, GoldenChain2}. Figure~\ref{fig:NegativityScaling} shows the ALN obtained from MPS simulations with bond dimension $\chi=500$ for the ground state of the golden chain for different system sizes $L$. Using the expected behavior of the LN in \eqref{eq:NegativityScaling}, we extract a central charge of $c=0.6975(8)$. This is very close to the expected value of $c=7/10$ and due to the fact that the central charge (of unitary CFTs) only takes certain discrete values for $c<1$~\cite{francesco2012conformal}, we can unambiguously associate our numerical result to the triciritical Ising CFT. 

The anyonic spin-$1$ model $\mathcal{H}_{\mathrm{spin-}1}(\theta)$ has three distinct gapless phases that feature $\mathbb{Z}_3$-, $\mathbb{Z}_5$- and $\mathbb{Z}_2$-sublattice symmetry and are described by CFTs with central charges $c=6/7$, $c=8/7$ and $c=6/7$, respectively~\cite{Spin1Models}. The corresponding values of $\theta$ that were used to extract the central charges via the ALN are $\theta=\pi /2$, $\theta=\pi$ and $\theta=3\pi /2$. Figure~\ref{fig:NegativityScaling} also shows the ALN for these three cases as obtained from MPS simulations with bond dimension $\chi=500$. The extracted central charges $c=0.8582(8)$, $c=1.1405(8)$ and $c=0.8507(6)$ are close to the expected values. In particular, we can again unambiguously associate the results for $\theta=\pi /2$ and $\theta=3\pi /2$ to CFTs with $c=6/7$ due to the discreteness of the central charge for $c< 1$.

Overall, our results show that the scaling behavior of the ALN can be used to extract the correct central charge of critical systems described by CFTs, which is in fact a nontrivial conclusion since the definition of the APT is more general than the ones of the BPT and FPT that may be considered limiting cases of the APT for which the scaling described in \eqref{eq:NegativityScaling} has aleady been demonstrated before~\cite{Calabrese_QFTnegativity, FPT_Shapourian_Ryu}.

\subsection{Topological Phase Transition}

\begin{figure}
\centering
\includegraphics{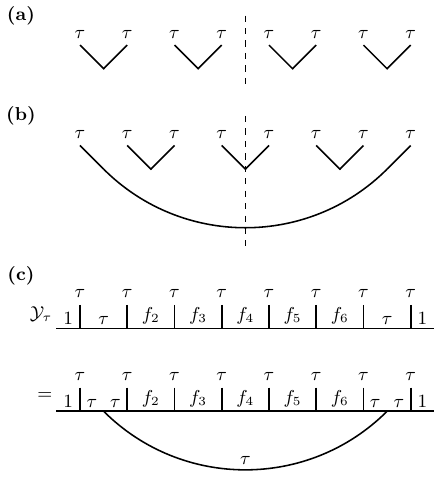}
\caption{Ground states of the dimerized golden chain $\mathcal{H}_{\mathrm{dGC}}(\Delta)$ for $L=8$ and \textbf{(a)} $\Delta = 1$, \textbf{(b)} $\Delta = -1$ consisting of dimers of charge $\tau$. For $\Delta = -1$, the dimerization between different unit cells leads to long-range entanglement between the two $\tau$s on the edges. The ALN for the bipartition indicated by the dashed lines is \textbf{(a)} $\mathcal{E}_a=0$, \textbf{(b)} $\mathcal{E}_a=2\ln(d_{\tau})$. \textbf{(c)} Action of $\mathcal{Y}_{\tau}$ on a fusion diagram for OBC with trivial boundary charges. A $\tau$ anyon tunnels between the first and the final unit cell, all other anyons in the fusion tree remain unaffected.}
\label{fig:FusionDiagramsDimerizedLimits}
\end{figure}

Let us now focus on gapped systems, for which the ALN can distinguish topologically nontrivial phases from trivial ones. Consider the golden chain Hamiltonian for Fibonacci anyons with an additional dimerization $\Delta$~\cite{PhysRevB.90.075129},
\begin{align}
	\mathcal{H}_{\mathrm{dGC}}(\Delta) = -\sum_{i=1}^{L-1} (1-(-1)^i\Delta) P_{i,i+1}^{(1)}.
	\label{eq:GoldenChainDimer}
\end{align}
Focusing on the case where the number of sites $L$ is even, we may interpret this Hamiltonian as having a two-site unit cell with intra-cell coupling $1+\Delta$ and inter-cell coupling $1-\Delta$ between neighboring $\tau$ anyons. In the limit $\Delta =1(\Delta =-1)$, only the intra-cell (inter-cell) coupling remains, which allows for an analysis of the ground state.\footnote{Here, we focus again on the case where the boundaries are trivial, for which the ground state is unique for both $\Delta = 1$ and $\Delta = -1$.} For $\Delta = 1$, the two anyons within each unit cell fuse to the trivial charge $1$ in the ground state, as depicted in Fig.~\ref{fig:FusionDiagramsDimerizedLimits}\textbf{(a)} for $L=8$, where the trivial charge labels $1$ have been omitted. For $\Delta = -1$, two neighboring $\tau$s of different unit cells fuse to $1$ such that there are two anyons left at the left and right boundary. These two anyons must also form a dimer with trivial fusion product, as shown in Fig.~\ref{fig:FusionDiagramsDimerizedLimits}\textbf{(b)} for $L=8$, since the total charge of the full system must be trivial. The reason for the latter constraint is the conservation of topological charge. When a physical system enters a topologically ordered phase, the total topological charge must be trivial due to the initial absence of topological charges, implying that arbitrarily exciting and braiding anyons cannot change this total charge. The major difference to the $\Delta = 1$ case is thus that the dimerization of the boundary anyons implies long-range entanglement. If we now choose to compute the ALN for the equal bipartition, we find $\mathcal{E}_a=0$ for $\Delta = 1$ and $\mathcal{E}_a=2\ln(d_{\tau})$ for $\Delta = -1$\footnote{Note that these values are only obtained when choosing the interface between the two subsystems to be in between two unit cells. Otherwise, $\mathcal{E}_a=\ln(d_{\tau})$ in both cases.} since we cut zero and two $\tau$-dimers, respectively (see Figs.~\ref{fig:FusionDiagramsDimerizedLimits}\textbf{(a)} and \ref{fig:FusionDiagramsDimerizedLimits}\textbf{(b)}), and the ALN of a single dimer of charge $c$ with trivial fusion product is $\ln(d_c)$. This interpretation can be confirmed by taking a look at the bond dimension $\chi$ of the respective MPS. For $\Delta=1$, we have $\chi=1$ at each bond, whereas for $\Delta=-1$, we have $\chi=2$ at bonds connecting two unit cells. Since a single dimer can be encoded by bond dimension $\chi=1$, it follows that there is a second dimer connecting two unit cells for $\Delta=-1$.

% For $\Delta = -1$, two neighboring $\tau$s of different unit cells fuse to $1$ such that there are two anyons left at the left and right boundary. Since the total charge of the system must be trivial, these two anyons must also form a dimer with trivial fusion product, as shown in Fig.~\ref{fig:FusionDiagramsDimerizedLimits}\textbf{(b)} for $L=8$. The major difference to the $\Delta = 1$ case is that this dimerization of the boundary anyons implies long-range entanglement.

We can go one step further and define an operator $\mathcal{Y}_{\tau}$ which introduces a $\tau$ anyon connecting the two ends of the fusion diagrams as depicted in Fig.~\ref{fig:FusionDiagramsDimerizedLimits}\textbf{(c)} for $L=8$. Note that at this point, the newly introduced anyon does not affect the other charges in the fusion tree. We can, however, bring this new fusion diagram back to its form before applying $\mathcal{Y}_{\tau}$ using the matrix elements
\begin{align}
\begin{split}
	&\langle f_1',f_2',\ldots f_{L-1}' |\mathcal{Y}_{\tau}| f_1,f_2,\ldots f_{L-1} \rangle\\
	&=\sqrt{d_{\tau}}[F^{\tau f_{2}\tau}_{f_1}]_{f_1f_2'}\prod_{i=3}^{L-1}[F^{\tau f_i \tau}_{f_{i-1}'}]_{f_{i-1}f_i'}.
\end{split}
\end{align}
Here, $f_i$ and $f_i'$ denote the fusion products with $f_1^{(\prime)}=f_{L-1}^{(\prime)}=\tau$ (see Fig.~\ref{fig:FusionDiagramsDimerizedLimits}\textbf{(c)}), $L$ the system size, $d_\tau$ again the quantum dimension of $\tau$ anyons and $F$ the $F$-moves. This operator may remind the reader of the topological symmetry operator $Y_\tau$, which is defined for periodic boundary conditions and can be interpreted as adding a closed $\tau$-loop encircling the anyonic chain~\cite{Spin1Models, Spin1Models2, GoldenChain}. We note that there are fundamental differences between $\mathcal{Y}_{\tau}$ as defined above and $Y_\tau$ that go beyond the boundary conditions. While $Y_\tau$ commutes with (the periodic version of) $\mathcal{H}_{\mathrm{dGC}}(\Delta)$, the same does not hold for $\mathcal{Y}_{\tau}$. Further, $Y_\tau^2=Y_\tau +1$, whereas $\mathcal{Y}_{\tau}^2=d_\tau^{-3/2}\mathcal{Y}_{\tau}+1$. Nevertheless, $\mathcal{Y}_{\tau}$ can be used to measure the presence of a long-range $\tau$-dimer between the two edges of the system. We denote the corresponding expectation value with respect to the ground state as $\mathcal{D}_{\tau} = \langle \mathcal{Y}_{\tau} \rangle$. Intuitively, when there is long-range entanglement in form of a $\tau$-dimer between the two edges, it is detected by $\mathcal{D}_{\tau}$ since $\mathcal{Y}_{\tau}$ adds another long-range $\tau$-dimer to the system. Combining these two dimers using $\tau \times \tau = 1+\tau$ yields a superposition of a long-range $\tau$-dimer and a trivial one. Thus, $\mathcal{D}_{\tau}\neq 0$ should take a finite value. On the other hand, if there is no long-range $\tau$-dimer, $\mathcal{Y}_{\tau}$ adds one ($1\times\tau =\tau$) to the system, resulting in a state that is orthogonal to the ground state. We therefore expect $\mathcal{D}_{\tau}=0$ in this case. We can verify this intuition for the two limiting cases and obtain $\mathcal{D}_{\tau}=1$ for $\Delta = -1$ and $\mathcal{D}_{\tau}=0$ for $\Delta =1$, in agreement with the above argument.% If the presence of the long-range $\tau$-dimer between the edges of the system is a property of the topological phase, we expect $\mathcal{D}_{\tau}$ to serve as a measure quantifying this nontrivial boundary for OBC.

\begin{figure}[t]
	\centering
	\includegraphics{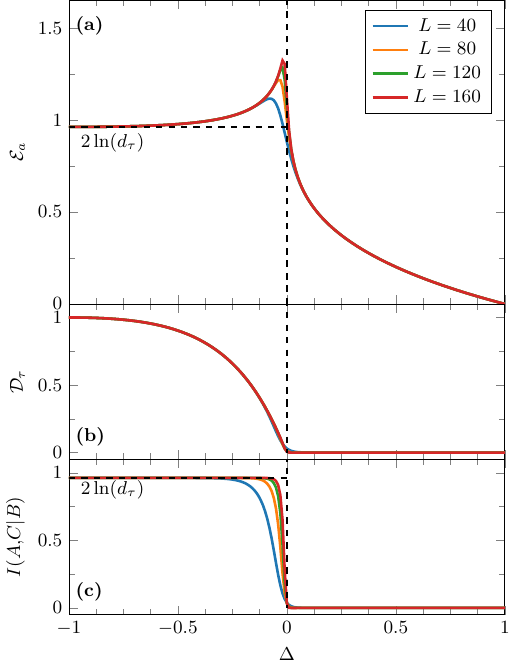}
	\caption{\textbf{(a)} ALN and \textbf{(b)} $\mathcal{D}_{\tau}$ of the ground state of the dimerized golden chain $\mathcal{H}_{\mathrm{dGC}}(\Delta)$ for dimerization $\Delta\in [-1,1]$ obtained from MPS simulations with bond dimension $\chi=300$. The bipartition is chosen such that the two subsystems are of equal size. $\Delta=0$ corresponds to the phase transition point; for $\Delta=1$($\Delta=-1$), the ALN agrees with $\mathcal{E}_a=0$($\mathcal{E}_a=2\ln(d_{\tau})$). \textbf{(c)} Conditional mutual information $I(A,C|B)$ for the partition depicted in Fig.~\ref{fig:CMI_partition}.}
\label{fig:PhaseTransitionDimer}
\end{figure}

%$\tau$ anyon tunneling from one edge to the other, it is detected by $\mathcal{Y}_{\tau}$ since $\tau \times \tau = 1+\tau$ such that the expectation value does not vanish. If there is no $\tau$ tunneling, we may view this as a tunneling process of the trivial charge $1$. Due to the fusion rule $1\times\tau =\tau$, this would intuitively lead to $\mathcal{O}=0$ since different tunneling charges correspond to orthogonal states. We can verify this intuition for the two limiting cases

Going beyond the above limiting cases, we show the ALN for $\Delta\in [-1,1]$ in Fig.~\ref{fig:PhaseTransitionDimer}\textbf{(a)} for system sizes up to $L=160$ and the corresponding values for $\mathcal{D}_{\tau}$ in Fig.~\ref{fig:PhaseTransitionDimer}\textbf{(b)}. These results were computed using MPS simulations with maximum bond dimension $\chi=300$. It can be seen that within both phases, the limiting case $|\Delta| = 1$ corresponds to the smallest entanglement. This can be explained by ``switching on'' the couplings that are trivial for $|\Delta| = 1$, leading to longer-ranged entanglement beyond two neigboring sites in the bulk, which is reflected accordingly in the ALN. For $\Delta=0$, $\mathcal{H}_{\mathrm{dGC}}(\Delta)$ becomes the critical golden chain model separating the topologically distinct phases~\cite{PhysRevB.90.075129}. It was already shown in Sec.~\ref{sec:NegativityScaling} that the ALN agrees with the CFT prediction at this point. In Fig.~\ref{fig:PhaseTransitionDimer}\textbf{(a)}, we can in addition see that the maximum of the ALN in the considered interval also seems to converge to the point $\Delta = 0$, showing that the ALN correctly captures the phase transition. Figure~\ref{fig:PhaseTransitionDimer}\textbf{(b)} shows that $\mathcal{D}_{\tau}=0$ in the trivial phase except for very small distances from the phase transition point $\Delta = 0$. Comparing the behavior for different system sizes suggests that this is merely a finite size effect such that $\mathcal{D}_{\tau}=0$ for $\Delta > 0$ in the thermodynamic limit. For $\Delta < 0$, finite values $\mathcal{D}_{\tau}\neq 0$ can be observed, saturating to $\mathcal{D}_{\tau}=1$ at $\Delta =-1$. Overall, we thus find that $\mathcal{D}_{\tau}$ can indeed be thought of as measuring the presence of a long-range $\tau$-dimer between the two edges of the system, which is, as expected, only nontrivial in the topological phase.

\begin{figure}[t]
\centering
\includegraphics{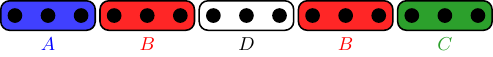}
\caption{Partition of a chain into four subsystems $A$, $B$, $C$ and $D$ used to compute the conditional mutual information $I(A,C|B)$.}
\label{fig:CMI_partition}
\end{figure}

Apart from $\mathcal{D}_{\tau}$, we can further confirm the topological nature of the $\Delta<0$ phase by directly studying the conditional mutual information $I(A,C|B)$, computed as~\cite{S_topo}
\begin{align}
	I(A,C|B) = S_{AB} + S_{BC} - S_B - S_{ABC},
	\label{eq:CMI}
\end{align}
for partitions of the system as indicated in Fig.~\ref{fig:CMI_partition}. Here, $S_X$ denotes the entanglement entropy obtained from the reduced density matrix associated with subsystem $X$. The results\footnote{For the computation of $I(A,C|B)$, the bond dimension $\chi =300$ of the previously obtained MPS was extended to $\chi =1000$. The need for a larger bond dimension is due to the fact that in order to compute the anyonic partial trace, the anyons to be traced out need to be exchanged to one of the edges of the system~\cite{bonderson_2007}, which was found to lead to much more entanglement. For $\chi =1000$, the truncation error when computing $I(A,C|B)$ was found to be at most $\mathcal{O}(10^{-3})$.} of $I(A,C|B)$ are shown in Fig.~\ref{fig:PhaseTransitionDimer}\textbf{(c)} for the same system sizes as before, where the partitions were chosen such that the sites are evenly distributed in the way shown in Fig.~\ref{fig:CMI_partition}. It can be seen that in the trivial phase, $I(A,C|B)$ converges to $0$, whereas in the topological phase, $I(A,C|B)$ converges to the entropy associated with two $\tau$-dimers, given by $2\ln(d_{\tau})$. This is expected since the long-range $\tau$-dimer between the edges of the system in the topological phase contributes $\ln(d_{\tau})$ both to $S_{AB}$ and $S_{BC}$, whereas it does not contribute to $S_B$ and $S_{ABC}$. The transition between these two converged values gets sharper and seems to converge to the critical point $\Delta = 0$ for larger system sizes.

Overall, we have demonstrated that the ALN can capture topological phase transitions and reflect the nontrivial entanglement between the boundaries of an open system. The presence of the nontrivial entanglement has been shown by explicitely evaluting the expectation value of a long-ranged dimer connecting the boundaries and by considering the conditional mutual information.

\section{Negativity Spectrum}
\label{sec:NegativitySpectrum}

Instead of only considering the ALN itself, one may expect to acquire further insights into the properties of the physical system by studying the eigenvalues of $\widetilde{\rho}^{T^a_A}$, named the negativity spectrum~\cite{negativity_Hamiltonian, NegSpectrumCFT, NegSpectrumGapped, NegSpectra_TwistedAndUntwisted, NegSpectrumRandomMixed}. The main motivation to consider this spectrum is that its analogue for the entanglement entropy, the entanglement sprectrum~\cite{EntanglementHamiltonian}, is known to feature universal information about the phase that goes beyond the entanglement entropy itself~\cite{EntanglementHamiltonian, EntanglementSpectrumExample1, EntanglementSpectrumExample2, EntanglementSpectrumExample3, EntanglementSpectrumExample4, EntanglementSpectrumExample5, EntanglementSpectrumExample6, EntanglementSpectrumExample7, EntanglementSpectrumExample8, EntanglementSpectrumExample9, EntanglementSpectrumExample10, EntanglementSpectrumExample11, EntanglementSpectrumExample12} (one should however be cautious when using the entanglement spectrum for such purposes~\cite{EntanglementSpectrumCaution}). We find below that for bipartitions, the APT possesses some nontrivial block structure that implies the presence of multiplets in the negativity spectrum which reflect the anyonic charges at the subsystems' interface and their exchange phases.

%Instead of only considering the ALN itself, one might be interested in the eigenvalues of $\widetilde{\rho}^{T^a_A}$, named the negativity spectrum~\cite{NegSpectrumCFT, NegSpectrumGapped, NegSpectra_TwistedAndUntwisted, NegSpectrumRandomMixed}, since it may contain further information. Indeed, we find below that for bipartitions, the APT possesses some nontrivial block structure that implies the presence of multiplets in the negativity spectrum.

\subsection{Block Structure of the Anyonic Partial Transpose}
\label{sec:BlockStructure}

%For the bipartition geometry, the APT possesses some nontrivial block structure. First, i
It can be seen from the fusion diagram in \eqref{eq:APT_sol} that there must be a block structure with respect to the total fusion product $g$, $\widetilde{\rho}^{T^a_A} = \oplus_g (\widetilde{\rho}^{T^a_A})_g$. This block structure is always present and does not depend on the geometry. If we focus on bipartitions, however, this block structure becomes much richer. Physically, the total topological charge of the full system must be trivial (i.e., $f=1$ in our notation), implying that $c_B=\overline{c}_A$ and $c'_B = \overbar{c'}\hspace{-2pt}_A$. It thus follows that
\begin{align}
	\widetilde{\rho}^{T^a_A} = \underset{(c,c';g), N_{cc'}^g>0}{\oplus} \left(\widetilde{\rho}^{T^a_A}\right)_{(c,c';g)},
	\label{eq:BlockStructure}
\end{align}
where $\left(\widetilde{\rho}^{T^a_A}\right)_{(c,c';g)}$ contains fusion diagrams with total fusion product $g$ and either $c_B=\overline{c}_A= c$, $c'_B = \overbar{c'}\hspace{-2pt}_A = c'$ or $c_B=\overline{c}_A= c'$, $c'_B = \overbar{c'}\hspace{-2pt}_A = c$. This means that $(c,c';g)$ and $(c',c;g)$ refer to the same block and thus only one of these two combinations should appear in the direct sum in \eqref{eq:BlockStructure}. From this block structure, it follows that the product $\widetilde{\rho}^{T^a_A}\left(\widetilde{\rho}^{T^a_A}\right)^\dagger$, which is needed to compute the ALN, also has some nontrivial structure:
\begin{align}
	\widetilde{\rho}^{T^a_A}\left(\widetilde{\rho}^{T^a_A}\right)^\dagger = \underset{(c,c';g), N_{cc'}^g>0}{\oplus} \Big[\widetilde{\rho}^{T^a_A}\left(\widetilde{\rho}^{T^a_A}\right)^\dagger \Big]_{(c,c';g)}.
	\label{eq:BlockStructure2}
\end{align}
Similar to $\widetilde{\rho}^{T^a_A}$, the product $\widetilde{\rho}^{T^a_A}\left(\widetilde{\rho}^{T^a_A}\right)^\dagger$ corresponds to a superposition of diagrams of the form depicted in \eqref{eq:APT_sol} with the crucial difference that the block $(c,c';g)$ is associated with the total fusion product $g$ and $\overline{c}_A = \overbar{c'}\hspace{-2pt}_A = c$, $c_B = c_B' = c'$. In this case, exchanging $c$ and $c'$ for $c\neq c'$ corresponds to going to a different block, unlike for $\widetilde{\rho}^{T^a_A}$ itself. That is, all combinations $(c,c';g)$ fulfilling the fusion consistency constraint $N_{cc'}^g>0$ appear in the direct sum in \eqref{eq:BlockStructure2}. It is this property that allows us to define the charge-resolved ALN for bipartitions in Sec.~\ref{sec:ChargeResolvedALN}.

\subsection{Structure of the Negativity Spectrum}
\label{sec:NegativitySpectrumStructure}

%Instead of simply looking at the LN itself, one might be interested in the eigenvalues of $\widetilde{\rho}^{T^a_A}$, named the negativity spectrum~\cite{NegSpectrumCFT, NegSpectrumGapped, NegSpectra_TwistedAndUntwisted, NegSpectrumRandomMixed}, since it may contain further information. Indeed, the block structure of $\widetilde{\rho}^{T^a_A}$ implies that for bipartitions, the negativity spectrum exhibits a rich structure. In the following, we focus on the bipartition geometry and thus set the total topological charge to the trivial charge.

%If the total charge is trivial, $f=1$ in \eqref{eq:APT_sol2} and it follows that $M(\mathbf{x},\mathbf{x'}; g) =  p\big(\mathbf{x},\mathbf{x'}; 1\big)R^{c_B'\overline{c}_A}_g/\sqrt{d_{c_B}d_{c_B'}}$. By choosing a gauge in which $R^{ab}_c=R^{ba}_c=\sqrt{\theta_c/(\theta_a\theta_b)}$, where $\theta_c$ denotes the topological twist of charge $c$~\cite{2102.05677}, and using the block structure (\eqref{eq:BlockStructure}), it follows that
%\begin{align}
%	\widetilde{\rho}^{T^a_A} = \underset{(c,c';g), N_{cc'}^g>0}{\oplus} R^{c'\overline{c}}_g T(c',\overline{c}),
%\end{align}
%with some hermitian matrix $T(c',\overline{c})$ (since $p\big(\mathbf{x},\mathbf{x'}; 1\big)$ must be hermitian) that is independent of $g$.\\

Consider again the bipartion geometry with trivial total charge $f=1$ and the general density matrix $\widetilde{\rho} = \sum_k p_k|\psi^{(k)} \rangle\langle \psi^{(k)} |$ with $\sum_k p_k = 1$. Each $|\psi^{(k)}\rangle$ can be written in terms of a Schmidt decomposition as $|\psi^{(k)}\rangle = \sum_{c,j_c}\Lambda^{(k)}_{c,j_c} |\phi^{A,(k)}_{c,j_c}\rangle|\phi^{B,(k)}_{c,j_c}\rangle$ with $\Lambda^{(k)}_{c,j_c}\in \mathbb{R}$, where $c$ labels the charge at the cut ($c_B$ in Fig.~\ref{fig:AnyonBasis}), $A,B$ the subsystem, $j_c$ the charge degeneracy and $|\phi^{A/B,(k)}_{c,j_c}\rangle$ the Schmidt states. We can now compute the APT for each of the contributions to $\widetilde{\rho}$ with respect to its respective Schmidt basis\footnote{The Schmidt basis does not necessarily agree with the basis choice in \eqref{eq:APT_def}. The result in \eqref{eq:APT_sol} can nevertheless be used since the coefficients only depend on the subsystems' total charges.}. By noting that the analogue to $p\big(\mathbf{x}, \mathbf{x'};f\big)$ in \eqref{eq:APT_def} is given by $\Lambda^{(k)}_{c,j_c}\Lambda^{(k)}_{c',j'_{c'}}$, the analogue to $M(\mathbf{x},\mathbf{x'}; g)$ in \eqref{eq:APT_sol} is $\Lambda^{(k)}_{c,j_c}\Lambda^{(k)}_{c',j'_{c'}}R^{c'\overline{c}}_g/\sqrt{d_cd_{c'}}$ since $f=1$.

By choosing a gauge in with $R^{ab}_c=R^{ba}_c=\sqrt{\theta_c/(\theta_a\theta_b)}$~\cite{2102.05677}, where $\theta_c$ denotes the topological twist of charge $c$, it can be seen that because the Schmidt values are real, the only complex phases contributing come from $R^{c'\overline{c}}_g$, which is identical for the full block $\left(\widetilde{\rho}^{T^a_A}\right)_{(c,c';g)}$. It thus follows that the APT of each contribution to $\widetilde{\rho}$ can be written as
\begin{align}
	\left(|\psi^{(k)} \rangle\langle \psi^{(k)} |\right)^{T^a_A} = \underset{(c,c';g), N_{cc'}^g>0}{\oplus} R^{c'\overline{c}}_g S(c',\overline{c}),
	\label{eq:NegSpec_intermediate}
\end{align}
where $S(c',\overline{c})$ is a real symmetric matrix independent of $g$. In particular, this implies that $S(c',\overline{c})$ and unitary transformations thereof are hermitian. This property allows us to write the APT of $\widetilde{\rho}$ as
\begin{align}
	\widetilde{\rho}^{T^a_A} = \underset{(c,c';g), N_{cc'}^g>0}{\oplus} R^{c'\overline{c}}_g T(c',\overline{c}),
	\label{eq:NegSpec_result}
\end{align}
with some hermitian matrix $T(c',\overline{c})$ that is independent of $g$. The reason for this is that we can use the result in \eqref{eq:NegSpec_intermediate} and transform each contribution $\left(|\psi^{(k)} \rangle\langle \psi^{(k)} |\right)^{T^a_A}$ to a common basis. Thus, $T(c',\overline{c})$ can be diagonalized and has real eigenvalues.

%\footnote{When taking a look at \eqref{eq:APT_sol}, it may be tempting to argue that $M(\mathbf{x},\mathbf{x'}; g) = p\big(\mathbf{x},\mathbf{x'}; 1\big) R^{c'\overline{c}}_g/\sqrt{d_{c_B}d_{c_B'}}$ for $f=1$ and use the hermiticity of $\widetilde{\rho}$, i.e., of $p$, to arrive at \eqref{eq:NegSpec_result}. This logic is incorrect since $p$ is hermitian with respect to exchanging $\mathbf{x}$ and $\mathbf{x'}$, whereas the basis elements of $\widetilde{\rho}^{T^a_A}$ depend on both.}.

Overall, we thus find the exciting result that the eigenvalues of $\widetilde{\rho}^{T^a_A}$, which we will refer to as $\lbrace z_i\rbrace$ from now on, lie in the complex plane on lines through the origin whose angles are specified by the exchange statistics given by $R^{c'\overline{c}}_g$. This means that from the negativity spectrum, we may identify which charges contribute the most to the entanglement. Further, the fact that $T(c',\overline{c})$ does not depend on $g$ implies exact degeneracies in the absolute values of the negativity spectrum $\lbrace z_i\rbrace$ if $N_{\overline{c}c'}^g>0$ for at least two different charges $g$. The different charges $g$ may however be distinguished by different $R$-moves $R^{c'\overline{c}}_g$. In particular, this implies that if such multiplets in the absolute values of the negativity spectrum exist, the anyons are necessarily non-abelian; if they are abelian, there is only a single $g$ with $N_{\overline{c}c'}^g>0$, i.e., the spectrum should not (generically) feature identical absolute values.

Note that the negativity spectrum may also be used to resolve the anyons' chirality. In this context, the precise definition of the APT in \eqref{eq:APT_def} is important. In principle, one can use an alternative definition where the direction of the braid is reversed. While this leaves the ALN and all Rényi negativities invariant, the eigenvalues of $\widetilde{\rho}^{T^a_A}$ become complex conjugated in the case discussed above, i.e., $R^{c'\overline{c}}_g\rightarrow (R^{c'\overline{c}}_g)^*$ in \eqref{eq:NegSpec_result} and thus $z_i\rightarrow z_i^*$.

\subsection{Example: Negativity Spectrum of Fibonacci Anyons}

\begin{figure}[t]
	\centering
	\includegraphics{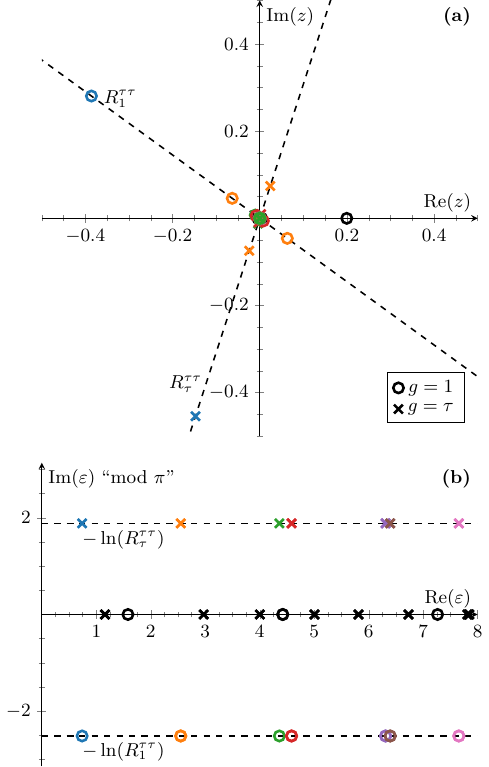}
	\caption{\textbf{(a)} Negativity spectrum $\lbrace z_i\rbrace$ for the golden chain for $L=17$ and a bipartition into $8$ and $9$ sites. Each multiplet $(c,c';g)=(\tau,\tau;1),(\tau,\tau;\tau)$ is highlighted by color, the fusion product $g$ is indicated by the marker. Eigenvalues not contained in a multiplet are displayed in black. \textbf{(b)} Logarithmic negativity spectrum $\lbrace \varepsilon_i\equiv -\ln(z_i) \rbrace$, which allows for easier verification of the multiplet structure. The imaginary values of $\varepsilon_i$ are changed by $\pm \pi$ if they do not agree with value of the corresponding $R$-move.}
	% Data: GS via ED
\label{fig:NegSpectrumGC}
\end{figure}

To illustrate these properties, we first study the golden chain model in \eqref{eq:GoldenChain} describing a chain of Fibonacci anyons once more. The nontrivial $R$-moves of these anyons are $R^{\tau\tau}_1=e^{4\pi i/5}$ and $R^{\tau\tau}_{\tau}=e^{-3\pi i/5}$. Figure~\ref{fig:NegSpectrumGC}\textbf{(a)} shows the negativity spectrum $\lbrace z_i\rbrace$ of the ground state for a bipartition into $8$ and $9$ sites for $L=17$ as obtained from exact diagonalization. Here, degeneracies in the absolute values are indicated by usage of the same color; the marker indicates the fusion product $g$. It can be seen that all $z_i$ either lie on the real axis, i.e., feature a trival $R$-move due to $c=1$ and/or $c'=1$, or on one of two other lines corresponding to multiplets $(c,c';g)=(\tau,\tau;1),(\tau,\tau;\tau)$.

Note that the color for all $z_i$ on the real axis is black to indicate that they do not belong to any multiplets that occur due to different charges $g$ as discussed in the context of \eqref{eq:NegSpec_result}. Nevertheless, there may of course be further degeneracies due to the nature of the \emph{state}, i.e., this is not a property of the negativity spectrum in the above sense. Such an additional degeneracy can also be observed for the orange multiplet in Fig.~\ref{fig:NegSpectrumGC}\textbf{(a)} containing four $z_i$, two for $g=1$ and two for $g=\tau$. The fact that $c=c'=\tau$ always ensures that the multiplet contains a multiple of two $z_i$.

For a better visualisation of the multiplet structure, Fig.~\ref{fig:NegSpectrumGC}\textbf{(b)} shows $\lbrace \varepsilon_i\equiv -\ln(z_i) \rbrace$, which is reminiscent of the entanglement spectrum of a density matrix~\cite{EntanglementHamiltonian}, i.e., $\widetilde{\rho}^{T^a_A}$ has eigenvalues $z_i=e^{-\varepsilon_i}$. Now, it is possible to easily verify the multiplet structre. Let us note that Fig.~\ref{fig:NegSpectrumGC}\textbf{(b)} does not necessarily show the correct imaginary part of $\varepsilon_i$. In order to avoid an unnessarily confusing picture, we added $\pm i\pi$ to $\varepsilon_i$ if its imaginary part does not match with the corresponding $R$-move. This is indicated by the label Im($\varepsilon$) ``mod $\pi$''.

\begin{figure}[t]
	\centering
	\includegraphics{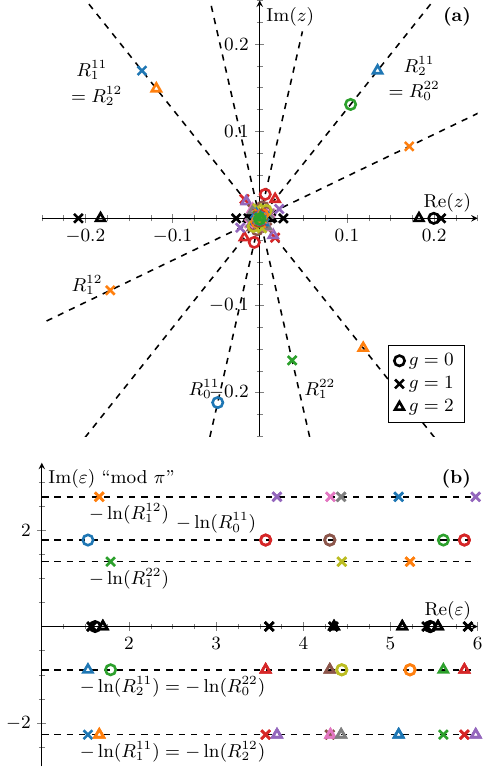}
	\caption{\textbf{(a)} Negativity spectrum $\lbrace z_i\rbrace$ for the $\mathrm{su}(2)_5$ anyon chain $\mathcal{H}_{\mathrm{spin-}1}(\theta=\pi/5)$ for $L=11$ and a bipartition into $5$ and $6$ sites. Each triplet $(c,c';g)=(1,1;0),(1,1;1),(1,1;2)$ and each of the doublets $(c,c';g)=(1,2;1),(1,2;2)$ and $(c,c';g)=(2,2;0),(2,2;1)$ are highlighted by color, the fusion product $g$ is indicated by the marker. Eigenvalues not contained in a multiplet are displayed in black. \textbf{(b)} Logarithmic negativity spectrum $\lbrace \varepsilon_i\equiv -\ln(z_i) \rbrace$. The imaginary values of $\varepsilon_i$ are changed by $\pm \pi$ if they do not agree with value of the corresponding $R$-move.}
	% Data: GS via ED
\label{fig:NegSpectrumSU2_5}
\end{figure}

\subsection{Example: Negativity Spectrum of \texorpdfstring{$\mathrm{su}(2)_5$}{su(2)5} Anyons}

Let us now consider the example given by $\mathcal{H}_{\mathrm{spin-}1}(\theta)$ in \eqref{eq:Spin1Model} for $\mathrm{su}(2)_5$ anyons. The nontrivial $R$-moves corresponding to exchanges of the integer spins $0,1,2$ are
\begin{align}
	&R^{11}_0= e^{-4\pi i/7}, \quad R^{11}_1=R^{12}_2=e^{5\pi i/7}, \quad R^{12}_1= e^{-6\pi i/7},\nonumber
	\\&R^{22}_1=e^{-3\pi i/7},   \quad R^{11}_2=R^{22}_0 =e^{2\pi i/7}. 
\end{align}
Figure~\ref{fig:NegSpectrumSU2_5} shows both the negativity spectrum $\lbrace z_i \rbrace$ and $\lbrace \varepsilon_i \rbrace$ for $\theta=\pi/5$, $L=11$ and a bipartion into $5$ and $6$ sites. It can be seen that the multiplet structure is much richer than for the Fibonacci anyon case: There are multiplets for $(c,c';g)=(1,1;0),(1,1;1),(1,1;2)$, $(c,c';g)=(1,2;1),(1,2;2)$ and $(c,c';g)=(2,2;0),(2,2;1)$, i.e., there are both triplets and doublets which can be distinguished from each other due to their unique combination of $R$-moves. In particular, we would even be able to identify these multiplets without the knowledge of $g$ by using the $R$-moves. All cases in which $c=0$ and/or $c'=0$ are due to their trivial exchange statistics again on the real axes and do not correspond to multiplets; the different combinations of $c$ and $c'$ may however still be distinguished using the fusion product $g$.

%Note that the negativity spectrum may be used to resolve the anyons' chirality. In this context, the precise definition of the APT in \eqref{eq:APT_def} is important. In principle, one can use an alternative definition where the direction of the braid is reversed. While this leaves the ALN and all Rényi negativities invariant, the eigenvalues of $\widetilde{\rho}^{T^a_A}$ become complex conjugated in the case discussed above, i.e., $R^{c'\overline{c}}_g\rightarrow (R^{c'\overline{c}}_g)^*$ in \eqref{eq:NegSpec_result} and thus $z_i\rightarrow z_i^*$ and $\varepsilon_i \rightarrow \varepsilon_i^*$.

\section{Charge- and Imbalance-Resolved Anyonic Logarithmic Entanglement Negativity}
\label{sec:ChargeImbalanceResolvedALN}
\subsection{Charge Resolution}
\label{sec:ChargeResolvedALN}

The block structure in \eqref{eq:BlockStructure2} (i.e., we assume again a bipartition with trivial total fusion product $f=1$) further allows for the definition of a charge-resolved ALN
\begin{align}
	\mathcal{E}_a(c) = \ln \sum_{c',g}\widetilde{\mathrm{Tr}}\sqrt{\Big[\widetilde{\rho}^{T^a_A}\left(\widetilde{\rho}^{T^a_A}\right)^\dagger \Big]_{(c,c';g)}},
	\label{eq:ChargeResolved}
\end{align}
which measures the entanglement between subsystems $A$ and $B$ mediated by anyons of charge $c$. Note that when there is no contribution to the density matrix $\widetilde{\rho}$ featuring charge $c$ at the cut/interface, $\mathcal{E}_a(c)=-\infty$. From the above definition, it is immediately clear that
\begin{align}
	\exp(\mathcal{E}_a) = \sum_c\exp\left(\mathcal{E}_a(c)\right),
\end{align}
as one would expect for a charge-resolved LN.

As an example, let us apply the notion of charge-resolved ALN to gapped systems. It was shown in Ref.~\cite{PhysRevB.99.115429} that in such systems, the entanglement spectrum features multiplets whose values (in the thermodynamic limit) can be related using the anyons' quantum dimensions. Knowing that the entanglement spectrum is related to the Schmidt decomposition, we can use this to estimate the charge-resolved ALN. Consider the ground state $|\psi\rangle$ of a gapped phase and its Schmidt decomposition $|\psi\rangle = \sum_c\sum_{j_c}\Lambda_{c,j_c}|\phi^A_{c,j_c}\rangle|\phi^B_{c,j_c}\rangle$, where $c$ is the anyonic charge at the interface between regions $A$ and $B$ and $j_c$ the corresponding degeneracy index, $|\phi^{A/B}_{c,j_c}\rangle$ denote the associated Schmidt states. The multiplet structure~\cite{PhysRevB.99.115429} dictates that $\Lambda_c^{\mathrm{multi}} \propto \sqrt{d_c}$, where $\Lambda_c^{\mathrm{multi}}$ denotes the Schmidt value associated with charge $c$ in a given multiplet. Note that not every charge $c$ needs to be contained in a multiplet; the degeneracy associated with $\Lambda_c^{\mathrm{multi}}$ is $N_{a_L\overbar{a}_R}^c$, with $a_L(\overbar{a}_R)$ being the charges at the left and right ends of subsystem $A$, and may thus become zero. Assuming that a single mutliplet $m$ is responsible for most of the entanglement, the charge-resolved ALN can be estimated to be
\begin{align}
	\mathcal{E}_a(c) \approx \ln\left( (\Lambda^{\mathrm{multi}})^2N_{a_L\overbar{a}_R}^c\sum_{c'\in m}\sum_g N_{a_L\overbar{a}_R}^{c'}N_{cc'}^gd_g\right).
	%\ln\left((\Lambda^{\mathrm{multi}})^2\right) + \ln\left( \sum_{c'\in m}\sum_g N_{a_L\overbar{a}_R}^cN_{a_L\overbar{a}_R}^{c'}N_{cc'}^gd_g\right).
	\label{eq:ChargeResolvedEstimate}
\end{align}
Here, $\Lambda^{\mathrm{multi}}$ is the renormalized Schmidt value associated with the dominant multiplet $m$, $\Lambda^{\mathrm{multi}}=\Lambda^{\mathrm{multi}}_c/\sqrt{d_c}$, where $c\in m$ can be any charge contained in the multiplet $m$.% Note that the second term in \eqref{eq:ChargeResolvedEstimate} can only contribute for non-abelian anyons.

Note that the multiplet structure discussed above implies that for the ground states of gapped systems, we expect the negativity spectrum to feature its usual (exact) multiplet structure as discussed in Sec.~\ref{sec:NegativitySpectrumStructure}, where multiple multiplets are at large but finite system sizes close to each other, merging to larger multiplets in the thermodynamic limit.

\subsection{Example: Charge Resolution for the Ground State of \texorpdfstring{$\mathcal{H}_{\mathrm{spin-}1}(\theta=0)$}{Hspin1(theta=0)} for \texorpdfstring{$\mathrm{su}(2)_5$}{su(2)5} Anyons}

Let us apply the notion of charge-resolved ALN to the ground state of the anyonic version of the AKLT Hamiltonian for $\mathrm{su}(2)_5$ anyons $\mathcal{H}_{\mathrm{spin-}1}(\theta=0)$. The ground state is known to feature only two contributions to the entanglement spectrum corresponding to a single multiplet with $\Lambda^{\mathrm{multi}} = d_2^{-1}$ in the thermodynamic limit~\cite{PhysRevB.99.115429}. These two contributions in the multiplet correspond to $c=0,1$ since $a_L=a_R=2$ and thus, we expect for the charge-resolved ALN in the thermodynamic limit
\begin{align}
	\mathcal{E}_a(c=0) &= \ln(d_2^{-2}(d_0+d_1))=\ln(d_0)=0,\\
	\mathcal{E}_a(c=1) &= \ln(d_2^{-2}(d_0+2d_1+d_2))=\ln(d_1).%\approx 0.809586916044713.
	\label{eq:ChargeResolvedPrediction}
\end{align}
Note that because there is only a single multiplet contributing to the entanglement for this specific state, the above predictions are actually exact for $L\rightarrow \infty$. This can be confirmed using matrix product states (MPS) for anyonic systems~\cite{AnyonMPS, AnyonMPS2, AnyonMPS3, AnyonMPS4}, where we found that the deviation from the values in \eqref{eq:ChargeResolvedPrediction} is $\mathcal{O}(10^{-4})$ for $L=20$ and an equal bipartition, reducing to $\mathcal{O}(10^{-9})$ for $L=50$ and to machine precision for $L=90$. Details on how to compute the (charge-resolved) ALN using MPS can be found in App.~\ref{app:MPS}. Interestingly, the above values for the charge-resolved ALN $\mathcal{E}_a(c)=\ln(d_c)$ agree with the ALN of a dimer of charge $c$ whose fusion outcome is the trivial charge~\cite{APT_Shapourian_Ryu}. We can thus interpret the contribution of charge $c=0,1$ to the entanglement as coming from an effective dimer of the respective charge between the two subsystems. This interpretation can be confirmed using the insights on the ground state discussed in Ref.~\cite{PhysRevB.99.115429}, which shows that there are two spin-2 connections between subsystems $A$ and $B$. Then, the total charge at the interface is $2\times 2=0+1$ and we expect the total entanglement to be identical to the one of two spin-2 dimers, which is indeed the case since $2\ln(d_2)=\ln(d_0)+\ln(d_1)$. The dimer interpretation is further supported by the MPS results showing that the bond dimension $\chi$ of the state is $\chi=2$, i.e., both $c=0$ and $c=1$ have a single Schmidt value contributing to the state, which effectively corresponds to a dimer of charge $c$ from an entanglement prespective. Overall, this interpretation confirms that the charged-resolved ALN is indeed charge-resolved.
%This interpretation does however neglect that the density matrix is a superposition which should intuitively result in rescaled dimer ALNs. The reason why this is not the case can be ultimately traced back to the nature of the APT as defined in \eqref{eq:APT_def}. The dimer interpretation is further supported by the MPS results showing that the bond dimension $\chi$ of the state is $\chi=2$, i.e., both $c=0$ and $c=1$ have a single Schmidt state contributing to the state, which effectively corresponds to a dimer from an entanglement prespective. Overall, this interpretation confirms that the charged-resolved ALN is indeed charge-resolved.

%Note that the multiplet structure discussed above implies that for the ground states of gapped systems, we expect the negativity spectrum to feature its usual (exact) multiplet structure as discussed in Sec.~\ref{sec:NegativitySpectrumStructure}, where multiple multiplets are at large but finite system sizes close to each other, merging to larger multiplets in the thermodynamic limit.

\subsection{Imbalance Resolution}
The possibilty of defining a charged-resolved ALN as in \eqref{eq:ChargeResolved} is itself highly nontrivial since charge resolution generally works differently for the LN. This can be seen for both bosonic and fermionic systems. While it is possible to define a symmetry-resolved entanglement entropy corresponding to the different symmetry sectors~\cite{Laflorencie_2014, PhysRevLett.120.200602}, this is not possible for the LN. The LN can however be resolved with respect to the associated charge \emph{imbalance} between the considered subsystems~\cite{ChargeResolvedNegativity, ChargeResolvedNegativity2}. It is thus a quite surprising result that it is possible to resolve the contributions of both the entanglement entropy and the ALN based on the anyonic charge at the interface between the two subsystems, which follows from the block structure in \eqref{eq:BlockStructure2}. Indeed, if we were to describe an abelian symmetry, such as particle number conservation, using the anyonic diagrammatics, we would recognise that the underlying assumption that $f=1$ made in Sec.~\ref{sec:BlockStructure} implies a certain total charge for \emph{each} contribution to the density matrix. Knowing this total charge implies that a charge imbalance resolution effectively becomes charge-resolved, just like the entanglement entropy. In this case, the block structure obtained without assuming $f=1$, using $\widetilde{\rho}^{T^a_A} = \oplus_g (\widetilde{\rho}^{T^a_A})_g$, corresponds to the charge imbalance-resolved LN without assuming that each contribution to the density matrix must have the same total charge. We therefore find that for abelian symmetries, the block structure leading to the nontrivial charge resolution is just a special case of the charge imbalanced resolution. In the case of non-abelian anyons however, this is not the case and we thus obtain a charge-resolved quantity for $f=1$ in addition to the charge imbalance-resolved ALN, that we may define as
\begin{align}
	\mathcal{E}_a^{\mathrm{imbalance}}(g) &= \ln \widetilde{\mathrm{Tr}}\sqrt{\left(\widetilde{\rho}^{T^a_A}\right)_g\left(\widetilde{\rho}^{T^a_A}\right)_g^{\dagger}}\label{eq:ChargeResolved_g}\\
	&\hspace{-3.75pt}\stackrel{f=1}{=} \ln \sum_{c,c'}\widetilde{\mathrm{Tr}}\sqrt{\Big[\widetilde{\rho}^{T^a_A}\left(\widetilde{\rho}^{T^a_A}\right)^\dagger \Big]_{(c,c';g)}}.
	\label{eq:ChargeResolved_g2}
\end{align}
Here, we do not rely on $f=1$ in order to define $\mathcal{E}_a^{\mathrm{imbalance}}(g)$. However, if $f=1$, we can make use of the richer block structure to express it as done in \eqref{eq:ChargeResolved_g2}. Assuming that we are dealing again with the ground state of a gapped Hamiltonian, we can estimate $\mathcal{E}_a^{\mathrm{imbalance}}(g)$ for $f=1$ as
\begin{align}
\begin{split}
	&\mathcal{E}_a^{\mathrm{imbalance}}(g) \\ 
	&\quad \approx\ln\left( (\Lambda^{\mathrm{multi}})^2d_g\sum_{c\in m}\sum_{c'\in m} N_{a_L\overbar{a}_R}^cN_{a_L\overbar{a}_R}^{c'}N_{cc'}^g\right).
\end{split}
	\label{eq:ChargeResolved_gEstimate}
\end{align}
Before applying this formula to the previous example, let us note that the imbalance-resolved ALN also fulfills the expected relation to the regular ALN,
\begin{align}
	\exp(\mathcal{E}_a) = \sum_g\exp\left(\mathcal{E}_a^{\mathrm{imbalance}}(g)\right).
\end{align}

\subsection{Example: Imbalance Resolution for the Ground State of \texorpdfstring{$\mathcal{H}_{\mathrm{spin-}1}(\theta=0)$}{Hspin1(theta=0)} for \texorpdfstring{$\mathrm{su}(2)_5$}{su(2)5} Anyons}

Going back to the ground state of $\mathcal{H}_{\mathrm{spin-}1}(\theta=0)$ for $\mathrm{su}(2)_5$ anyons, we use \eqref{eq:ChargeResolved_gEstimate} to compute the imbalance-resolved ALN in the thermodynamic limit
\begin{align}
	\mathcal{E}_a^{\mathrm{imbalance}}(g=0) &= \ln(2d_2^{-2}),\\
	\mathcal{E}_a^{\mathrm{imbalance}}(g=1) &= \ln(3d_1d_2^{-2}),\\
	\mathcal{E}_a^{\mathrm{imbalance}}(g=2) &= \ln(d_2^{-1}).
	\label{eq:ChargeResolved_gPrediction}
\end{align}
These values have been verified with approximately the same deviation as for $\mathcal{E}_a(c)$ using MPS. Note that the above nontrivial values of the imbalance-resolved ALN do not contradict the interpretation that the state effectively corresponds to two dimers since the imbalance is considered for the partially transposed density matrix rather than the state itself. In fact, it is clear from the discussion of the negativity spectrum in Sec.~\ref{sec:NegativitySpectrumStructure} that due to the possibility $c=1$, the imbalance-resolved ALN is for $g=0,1,2$ nontrivial. The interpretation of $\mathcal{E}_a^{\mathrm{imbalance}}(g)$ is generally harder than that of $\mathcal{E}_a(c)$. In the above case, we know that the charge blocks of APT are specified by $c=0,1$, $c'=0,1$ and all consistent values of $g$. The $g=1$ imbalance sector contributes the most to the entanglement since the three sectors $(c,c')=(0,1),(1,0),(1,1)$ are of importance. On the other hand, for $g=0$, two sectors $(c,c')=(0,0),(1,1)$ are relevant and for $g=2$, only $(c,c')=(1,1)$ contributes. This gives a rough intuition why $\mathcal{E}_a^{\mathrm{imbalance}}(g=1)>\mathcal{E}_a^{\mathrm{imbalance}}(g=0)>\mathcal{E}_a^{\mathrm{imbalance}}(g=2)$ in our case.

\section{Conclusion}
\label{sec:Conclusion}

We studied the generalization of the partial transpose and the corresponding entanglement measure, the logarithmic negativity (LN), to anyonic systems and revealed some fundamental properties of these generalizations. We showed that when applied to fermionic systems, the anyonic partial transpose (APT) reproduces the fermionic partial transpose up to a unitary transformation, implying that the APT is indeed an appropriate generalization of the partial transpose to anyonic systems. Focusing on low-energy properties, we numerically verified that the anyonic logarithmic negativity (ALN) of the ground state of a system described by CFT exhibits the expected scaling behavior. For gapped systems, it was found that the ALN captures the phase transition between a topologically nontrivial and a trivial phase. For bipartitions with trivial total charge, the APT possesses some rich block structure, which leads to the negativity spectrum consisting of multiplets whose constituents feature identical absolute values and complex phases that are related to the anyonic exchange statistics. In the presence of the aforementioned block structure, we further discovered the possibility of studying the charge-resolved ALN, which is analogous to the charge resolution of the entanglement entropy~\cite{Laflorencie_2014, PhysRevLett.120.200602}. This goes beyond what has been studied for bosonic and fermionic systems, for which, on the other hand, a resolution with respect to the charge imbalance was found~\cite{ChargeResolvedNegativity, ChargeResolvedNegativity2}. Such a charge imbalance-resolved LN can also be studied for anyonic systems, which does not rely on the block struture required for charge resolution.

Given the above results, there are many interesting questions one may consider for future research. On the level of pure applications, one may aim to apply the charge-resolved ALN to analyze bipartitions of mixed states corresponding to, e.g., thermal density matrices at finite temperatures for three-dimensional topologically ordered systems. In this context, we have to note that the notion of charge resolution can also be applied to spin systems, i.e., it should be possible to resolve the LN with respect to the total spin of the subsystems. It would be very interesting to see whether additional insights can be provided using this charge resolution.

Going beyond the bipartite geometry, we can no longer resolve the ALN with respect to the charges of the subsystems. Suppose we divide the full system into subsystems $A$, $B$ and $C$. To find the entanglement between $A$ and $B$, we need to partially trace the density matrix over subsystem $C$ before computing the PT such that the total charge after the partial trace is in general no longer trivial. However, it should be possible to still resolve the ALN with respect to the charges of subsystem $A(B)$ by promoting the total charge of the partially traced density matrix to some ``auxiliary charge'' that is now part of the other subsystem $B(A)$. This promotion does not affect the total charge of subsystem $A(B)$ while ensuring that the total charge is trivial. This means that we can now resolve the ALN with respect to the total charge of $A(B)$, which intuitively seems to be a sensical extension. Working out the details and insights that can be learned from studying such a construction looks like an interesting and promising future direction. In particular, in this case, one may be able to see nontrivial differences when resolving with respect to the charge of subsystem $A$ compared to the resolution with respect to subsystem $B$.

\section*{Acknowledgements}

The authors thank Adam Smith for helpful discussions. This work was supported by the European Research Council (ERC) under the European Union’s Horizon 2020 research and innovation program under Grant Agreement No. 771537 and No. 851161, the Deutsche Forschungsgemeinschaft (DFG, German Research Foundation) under Germany’s Excellence Strategy EXC-2111-390814868, TRR 360 - 492547816, and the Munich Quantum Valley, which is supported by the Bavarian state government with funds from the Hightech Agenda Bayern Plus.

The programming effort was partially reduced by using the TeNPy library~\cite{Tenpy} as template for some of the MPS algorithms.

\section*{Data Availability}

Access to the numerical data and simulation codes may be granted upon reasonable request~\cite{zenodo}.

\appendix

\section{Computation of the Anyonic Partial Transpose for an Anyon Dimer}
\label{app:APTDimer}

In this section, we show how \eqref{eq:APT_sol} can be derived using the standard operations on fusion diagrams that are introduced, e.g., in Refs.~\cite{APT_Shapourian_Ryu, simon2020topological, 1506.05805, 0707.4206, bonderson_2007, 2102.05677, InsideOutsideBases, PhysRevB.107.195129}. This can be done by considering the case of an anyonic dimer since the corresponding result straight forwardly generalizes to \eqref{eq:APT_sol} due the choice of basis for the fusion diagrams. The calculation for anyon dimers is
\begin{widetext}
\begin{align}
\begin{split}
	&\frac{1}{N_f}\left(\hspace{-12.5pt}\vcenter{\includegraphics{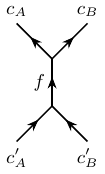}}\hspace{-453pt}\right)^{T_A^a} \stackrel{(1)}{=}\frac{1}{N_f} \hspace{-5pt}\vcenter{\includegraphics{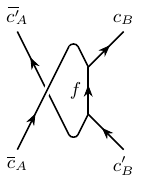 }}\hspace{-433pt} \stackrel{(2)}{=}\frac{1}{N_f} \hspace{-8pt}\vcenter{\includegraphics{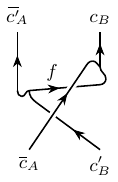 }}\hspace{-438pt} \stackrel{(3)}{=}\frac{1}{N_f} A^{c_Ac_B}_f \Big(A^{c_A'c_B'}_f\Big)^* \hspace{-8pt} \vcenter{\includegraphics{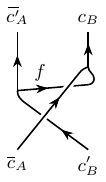 }}\hspace{-438pt}\\
	&\stackrel{(4)}{=}\frac{1}{N_f} \sum_g \sqrt{\frac{d_g}{d_{c_A}d_{c'_B}}} A^{c_Ac_B}_f \Big(A^{c_A'c_B'}_f\Big)^* R_g^{\overline{c}_Ac'_B}\Big(R_{c_B}^{f\overline{c}_A}\Big)^*\hspace{-8pt} \vcenter{\includegraphics{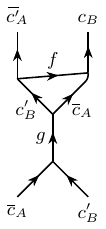 }}\hspace{-438pt}\\
	&\stackrel{(5)}{=} \sum_g \frac{\sqrt{d_fd_g}}{N_f}\frac{1}{\sqrt{d_{c_B}d_{c'_B}}} A^{c_Ac_B}_f \Big(A^{c_A'c_B'}_f\Big)^* R_g^{\overline{c}_Ac'_B}\Big(R_{c_B}^{f\overline{c}_A}\Big)^*\Big[F_g^{\overbar{c'}_{\hspace{-2pt}A}f\overline{c}_A}\Big]_{c_B'c_B} \hspace{-8pt} \vcenter{\includegraphics{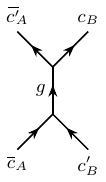 }}\hspace{-448pt},
\end{split}
\end{align}
\end{widetext}
where after $(1)$ performing the partial transpose, we first $(2)$ deform the fusion diagram and then $(3)$ apply $A$-moves. We then $(4)$ insert a resolution of identity twice at the crossing of $\overline{c}_A$ with $c'_B$ and $f$, respectively, and apply the corresponding $R$-moves. After $(5)$ applying a $F$-move, one arrives at the final result. In the above equation, $N_f$ again refers to the normalization given in \eqref{eq:APT_normalization}. Further, $\sqrt{d_fd_g}/N_f = d_g/N_g$, which confirms the prefactors in \eqref{eq:APT_sol}. Note that the above result has already been derived for the case $c_A=c'_A$, $c_B=c'_B$ and higher fusion multiplicities in Ref.~\cite{APT_Shapourian_Ryu}.

\section{Proof of the Relation of the Fermionic Partial Transpose to the Anyonic Partial Transpose Applied to Fermionic Systems}
\label{app:FPTandAPT}

In the following, we show that the FPT can indeed be related to the APT using the relations in Sec.~\ref{sec:RelationAPT-FPT}. First of all, note that when describing fermions using the anyon diagrammatics, all charges corresponding to $a_i^{(\prime)},  1\leq i \leq N$ and $b_j^{(\prime)}, 1\leq j \leq M$ in \eqref{eq:APT_def} are fermionic, denoted by $\psi$ with the $R$-move $R^{\psi\psi}_1=-1$ encoding the exchange statistics (where $1$ again denotes the trivial charge). Since the $F$-moves for the fermionic anyon model are trivial, i.e., only enforce fusion consistency without containing any nontrivial phases, it can already be seen from \eqref{eq:APT_sol2} that the APT (in an appropriate basis in which the initial density matrix is real symmetric) can only have real entries, implying a real negativity spectrum. Thus, it can only be connected to the twisted FPT, which also possesses a real spectrum, whereas the usual FPT possesses a complex one~\cite{NegSpectra_TwistedAndUntwisted}. Since in the definition of the APT in \eqref{eq:APT_def}, the order of the anyons in subsystem $A$ is reversed, it is natrual to apply the spatial reflection operator $\mathcal{R}$ defined by $\mathcal{R} f_j \mathcal{R}^{-1} = if_{L+1-j}$~\cite{FermionReflectionOperator} onto subsystem $A$. It turns out that this is already enough to make the APT and the twisted FPT agree with each other, i.e.,
\begin{align}
	\widetilde{\rho}^{T^a_A} = \mathcal{R}_A\rho^{T^f_A}(-1)^{F_A}\mathcal{R}_A^{-1},
	\label{eq:ConnectionAPT-FPT_app}
\end{align}
where $F_A = \sum_{j\in A}n_j$. We will show this relation in the following by considering how the operators on the right-hand side of \eqref{eq:ConnectionAPT-FPT_app} act in the fermionic occupation basis:
\begin{widetext}
\begin{align}
\begin{split}
	&\mathcal{R}_A\left(|\lbrace n_j\rbrace_A,\lbrace n_j\rbrace_B \rangle  \langle\lbrace \overbar{n}_j\rbrace_A, \lbrace\overbar{n}_j\rbrace_B|\right)^{T_A^f}(-1)^{F_A}\mathcal{R}_A^{-1}\\
	%= (-1)^{\phi(\lbrace n_j\rbrace,\lbrace \overbar{n}_j\rbrace)} \mathcal{R}_A |\lbrace \overbar{n}_j\rbrace_A,\lbrace n_j\rbrace_B \rangle  \langle\lbrace n_j\rbrace_A, \lbrace\overbar{n}_j\rbrace_B|(-1)^{F_A}\mathcal{R}_A^{-1}\\
	&= (-1)^{\phi(\lbrace n_j\rbrace,\lbrace \overbar{n}_j\rbrace)+\tau_A} \mathcal{R}_A |\lbrace \overbar{n}_j\rbrace_A,\lbrace n_j\rbrace_B \rangle  \langle\lbrace n_j\rbrace_A, \lbrace\overbar{n}_j\rbrace_B|\mathcal{R}_A^{-1}\\
	&= (-1)^{\phi(\lbrace n_j\rbrace,\lbrace \overbar{n}_j\rbrace)+\tau_A}i^{\tau_A}(-i)^{\overbar{\tau}_A}(-1)^{((\tau_A-1)\tau_A + (\overbar{\tau}_A-1)\overbar{\tau}_A)/2} |\lbrace \overbar{n}_j\rbrace_A^{\mathcal{R}},\lbrace n_j\rbrace_B \rangle  \langle\lbrace n_j\rbrace_A^{\mathcal{R}}, \lbrace\overbar{n}_j\rbrace_B|,
\end{split}
	\label{eq:TrafoFPT}
\end{align}
\end{widetext}
where $\phi(\lbrace n_j\rbrace,\lbrace \overbar{n}_j\rbrace)$ is given in \eqref{eq:FermionPhaseFactor}. Here, the same notation as in \eqref{eq:FPT} has been used, with the addition of the superscript $\mathcal{R}$ in $\lbrace n_j\rbrace_A^{\mathcal{R}}$, which indicates that the particle configuration is spatially reflected. In \eqref{eq:TrafoFPT}, the factor of $(-1)^{((\tau_A-1)\tau_A + (\overbar{\tau}_A-1)\overbar{\tau}_A)/2}$ arises from the exchanges of fermionic creation and annihilation operators due to the spatial reflection. Since the above states is exactly what is also obtained from the APT, we can now compare the phase to the one obtained when applying the APT, which can be expressed as $(-1)^{\tau_A\overbar{\tau}_B}(-1)^{\tau_A({\tau}_A+{\tau}_B)} = (-1)^{\tau_A(\overbar{\tau}_B+\tau_A+\tau_B)}$. The relative phase between the FPT and the APT is thus
\begin{widetext}
\begin{align}
%\begin{split}
	&(-1)^{\tau_A(\tau_A+2)/2+\overbar{\tau}_A(\overbar{\tau}_A+2)/2+\tau_B\overbar{\tau}_B + \tau_A\tau_B + \overbar{\tau}_A\overbar{\tau}_B + (\tau_A + \tau_B)(\overbar{\tau}_A + \overbar{\tau}_B)+3\tau_A/2+3\overbar{\tau}_A/2 - \tau_A(\tau_A-1)/2-\overbar{\tau}_A(\overbar{\tau}_A-1)/2 + \tau_A(\overbar{\tau}_B+\tau_A+\tau_B)}\nonumber\\
	%&= (-1)^{3\tau_A+3\overbar{\tau}_A+\tau_B\overbar{\tau}_B+ 2\tau_A\tau_B + \overbar{\tau}_A\overbar{\tau}_B + (\tau_A + \tau_B)(\overbar{\tau}_A + \overbar{\tau}_B) + \tau_A(\overbar{\tau}_B+\tau_A)}
	%= (-1)^{\tau_A+\overbar{\tau}_A + \overbar{\tau}_A\overbar{\tau}_B + (\tau_A + \tau_B)(\overbar{\tau}_A + 2\overbar{\tau}_B) + \tau_A^2}\\
	%&= (-1)^{\tau_A(\tau_A+1) + \overbar{\tau}_A(\tau_A + \tau_B+\overbar{\tau}_B+1)}
	&= (-1)^{ \overbar{\tau}_A(\tau_A + \tau_B+\overbar{\tau}_B+1)}.
%\end{split}
\end{align}
\end{widetext}
It is straight forward to verify that this relative phase is trivial if $(\tau_A+\tau_B)\mathrm{mod} 2 = (\overbar{\tau}_A+\overbar{\tau}_B)\mathrm{mod} 2$, which expresses the conservation of fermion parity, i.e., this relation must always hold in physical systems. With this, it is shown that \eqref{eq:ConnectionAPT-FPT_app} holds.

Let us now turn to the second connection between the APT and the FPT, given by
\begin{align}
	\widetilde{\rho}_{\sigma}^{T^a_A} = \frac{\theta_{\sigma}}{d_{\sigma}} U \mathcal{R}_A\rho^{T^f_A}\mathcal{R}_A^{-1}U^{\dagger}.
	\label{eq:ConnectionAPT-FPT2_app}
\end{align}
with
\begin{align}
	U \left|\lbrace n_j\rbrace_A,\lbrace n_j\rbrace_B \right\rangle = (-i)^{\tau_A \mathrm{mod}2} \left|\lbrace n_j\rbrace_A,\lbrace n_j\rbrace_B \right\rangle.
\end{align}
First of all, note that $\widetilde{\rho}_{\sigma}$ is an anyonic density matrix representing the fermionic system but has in addition a Majorana fermion $\sigma$ as boundary charge in subsystem $A$. Concretely, this means that in the basis in Fig.~\ref{fig:AnyonBasis}, there is an additional anyon $a_0=\sigma$ that does not affect the fermionic system described by the density matrix in any other sense. Further, there is no operator acting on the fermionic density matrix that can create such a boundary charge, i.e., we either assume that it is already there to begin with and does not affect the FPT or we simply identify states with identical fermion configuration and different boundary charges with each other. Since the origin of the phases and operations in \eqref{eq:ConnectionAPT-FPT2_app} has already been discussed in Sec.~\ref{sec:RelationAPT-FPT}, we directly go to the computation of the phases that arise on both sides of \eqref{eq:ConnectionAPT-FPT2_app}. Before doing so, however, we need to specify how to describe the Majorana fermions in terms of anyon models. We choose the Ising anyon model~\cite{APT_Shapourian_Ryu, bonderson_2007} to do so. The nontrivial fusion processes are $\sigma\times\sigma=1+\psi$, $\sigma\times\psi=\sigma$ and $\psi\times\psi =1$, where $\psi$ again denotes the fermion and $1$ the trivial charge and each anyon is their own anticharge. The corresponding $R$-moves are 
\begin{align}
	R^{\sigma\sigma}_1=e^{-i\frac{\pi}{8}},\quad\,\,\, R^{\sigma\sigma}_{\psi}=e^{i\frac{3\pi}{8}},\\
	R^{\sigma\psi}_{\sigma}=R^{\psi\sigma}_{\sigma}=-i,\quad R^{\psi\psi}_1=-1,
\end{align}
the nontrivial $F$-moves are
\begin{align}
	&\big[F^{\psi\sigma\psi}_{\sigma}\big]_{\sigma\sigma}=\big[F^{\sigma\psi\sigma}_{\psi}\big]_{\sigma\sigma}=-1,\,\,\,\, \big[F^{\sigma\sigma\sigma}_{\sigma}\big]_{\psi\psi}=-\frac{1}{\sqrt{2}},\\
	&\big[F^{\sigma\sigma\sigma}_{\sigma}\big]_{11}=\big[F^{\sigma\sigma\sigma}_{\sigma}\big]_{1 \psi}=\big[F^{\sigma\sigma\sigma}_{\sigma}\big]_{\psi 1}=\frac{1}{\sqrt{2}}.
\end{align}
We can now determine the phases arising from \eqref{eq:APT_sol2} to be
\begin{align}
	\big(R^{\psi\sigma}_{\sigma}\big)^{\overbar{\tau}_B\mathrm{mod}2}\big(R^{\sigma\sigma}_1\big)^*e^{-i\frac{\pi}{2}(\tau_B\mathrm{mod}2)}(-1)^{\tau_B\overbar{\tau}_B}.
\end{align}
Here, it was used that no matter the number of fermions in subsystem $A$, $c_A=c_A'=\sigma$ due to the additonal boundary charge. The remaining effect of $c_B^{(\prime)}=1,\psi$ is encoded in the $\mathrm{mod}2$ terms. On the other hand, the phases arising from the FPT and the additional transformations in \eqref{eq:ConnectionAPT-FPT2_app} are
\begin{align}
	&(-1)^{\phi(\lbrace n_j\rbrace,\lbrace \overbar{n}_j\rbrace)}i^{\tau_A}(-i)^{\overbar{\tau}_A}(-1)^{((\tau_A-1)\tau_A + (\overbar{\tau}_A-1)\overbar{\tau}_A)/2}\nonumber\\
	&\quad\times \theta_{\sigma} i^{\tau_A \mathrm{mod}2} (-i)^{\overbar{\tau}_A \mathrm{mod}2}
\end{align}
Since we already know from the previous consideration that the transformed states themselves agree (up to the identification that ignores the boundary charge), we can focus again on the relative phase, where we use $\theta_{\sigma}=(R^{\sigma\sigma}_1)^*$:%\textcolor{red}{[some intermediate steps will be deleted later on; I let them here to make it easier to verify the calculation]}
\begin{widetext}
\begin{align}
\begin{split}
	& i^{\tau_B \mathrm{mod}2 + \overbar{\tau}_B \mathrm{mod}2}(-1)^{\tau_B\overbar{\tau}_B}  (-1)^{\phi(\lbrace n_j\rbrace,\lbrace \overbar{n}_j\rbrace)}i^{\tau_A + \tau_A \mathrm{mod}2}(-i)^{\overbar{\tau}_A+\overbar{\tau}_A \mathrm{mod}2}(-1)^{((\tau_A-1)\tau_A + (\overbar{\tau}_A-1)\overbar{\tau}_A)/2}\\
	&= (-1)^{\tau_A(\tau_B+1) + \overbar{\tau}_A(\overbar{\tau}_B+1)+(\tau_A + \tau_B)(\overbar{\tau}_A + \overbar{\tau}_B)+(-\tau_A \mathrm{mod}2+\tau_B \mathrm{mod}2-\overbar{\tau}_A \mathrm{mod}2+\overbar{\tau}_B \mathrm{mod}2)/2}.
\end{split}
\end{align}
\end{widetext}
It can be verified that this relative phase is indeed trivial if $(\tau_A+\tau_B)\mathrm{mod} 2 = (\overbar{\tau}_A+\overbar{\tau}_B)\mathrm{mod} 2$ by considering all consistent combinations of even and odd $\tau_{A,B},\overbar{\tau}_{A,B}$. With this, we conclude that the correspondence between the APT and the FPT in \eqref{eq:ConnectionAPT-FPT2_app} holds.

Since the APT applied to bosonic systems is similarly related to the bosonic PT via spatial reflection of region $A$, it is justified to refer to the APT as the generalization of both the bosonic and the fermionic PT to anyonic systems.

\section{Computing the Anyonic Entanglement Negativity with Matrix Product States}
\label{app:MPS}

\begin{figure*}[t]
\centering
\includegraphics{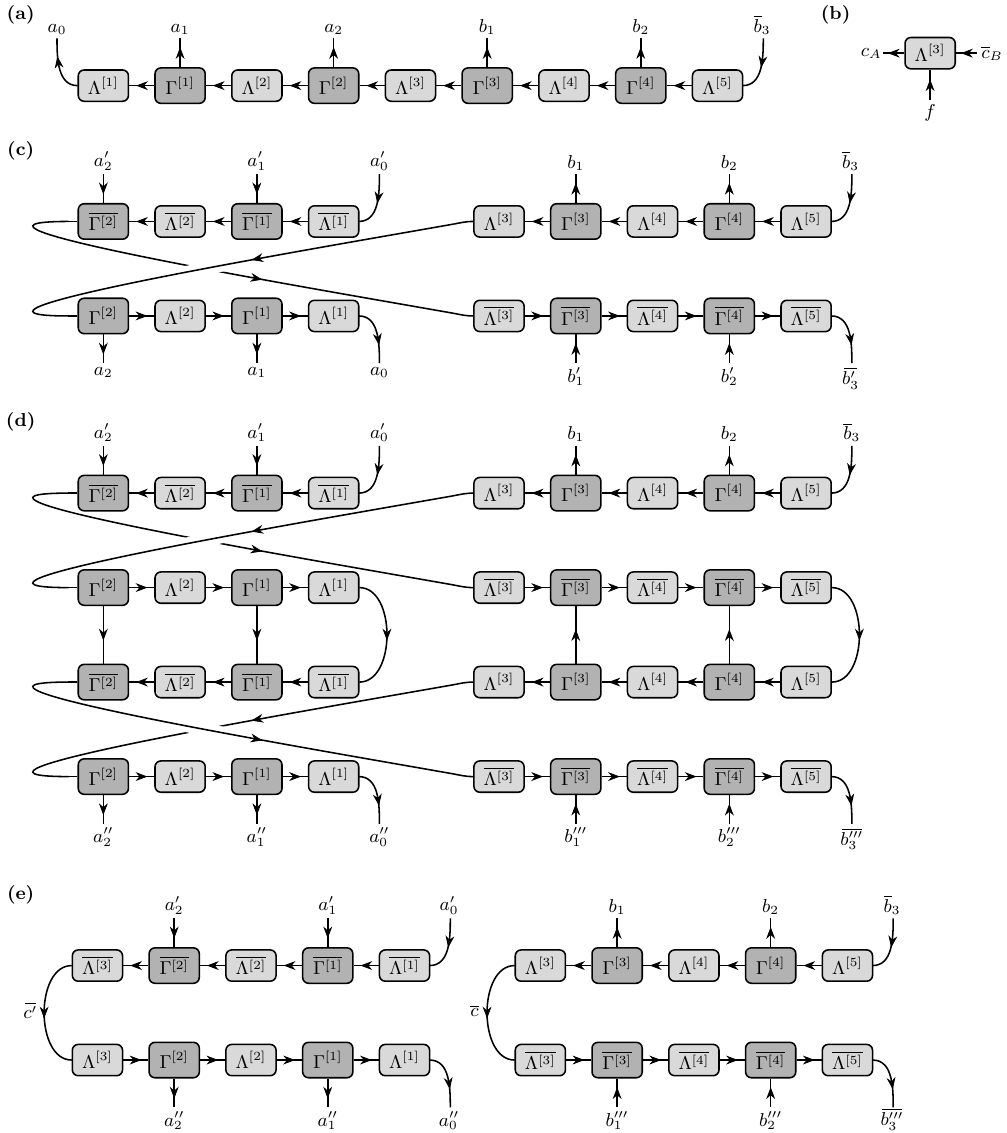}
\caption{\textbf{(a)} MPS representing a state in the basis in Fig.~\ref{fig:AnyonBasis} for $N=M=2$ and $f=1$ in canonical form, where the labels of the fusion products are omitted for convenience. The anyonic charges at the boundary of the MPS are promoted to additional (boundary) charges $a_0$ and $\overline{b}_3$ belonging to subsystem $A$ and $B$, respectively. \textbf{(b)} In the case $f\neq 1$, the state can be represented by using a 3-leg tensor at the interface between the two subsystems instead of the usual $\Lambda^{[i]}$ tensor; for the case considered in \textbf{(a)}, $\Lambda^{[3]}$ would need to have this additional leg. \textbf{(c)} APT $\widetilde{\rho}^{T_A^a}$ for the MPS in \textbf{(a)}. \textbf{(d)} $\left(\widetilde{\rho}^{T_A^a}\right)^\dagger\widetilde{\rho}^{T_A^a}$ using the APT in \textbf{(c)}. By exploiting the orthogonality properties of the canonical form, this operator can be simplified to the operator shown in \textbf{(e)}. Here, we added the labels of the charges $c$ and $c'$ dictating the block structure \eqref{eq:BlockStructure2} (the corresponding charges $g$ are obtained by fusing $c$ and $c'$.)}
\label{fig:MPS}
\end{figure*}

In this section, we explain how to compute the ALN and its charge- and imbalance-resolved variants using MPS. For this, we assume the reader to be acquainted with the fundamentals of tensor networks~\cite{MPS2, MPS3, MPS4, MPS1, Tenpy}. For the generalization to anyonic systems~\cite{AnyonMPS, AnyonMPS2, AnyonMPS3, AnyonMPS4, AnyonMERA}, we add (directed) anyonic charges to the tensor legs which obey the anyonic fusion rules and with respect to which all tensors are block diagonal. An example is given in Fig.~\ref{fig:MPS}\textbf{(a)}, which corresponds to a MPS representing a state in the basis given in Fig.~\ref{fig:AnyonBasis} for $N=M=2$ and $f=1$ in canonical form. We denote the canonical 3-leg tensors as $\Gamma^{[i]}$ and the (diagonal) 2-leg tensors containing the Schmidt values as $\Lambda^{[i]}$. In Fig.~\ref{fig:MPS}, we omitted the labels of the intermediate fusion products for convenience and bent the lines at the boundaries associated with $a_0$ and $\overline{b}_3$ upwards, that is, we effectively promote the corresponding charges to be part of subsystems $A$ and $B$, respectively. In practice, this does not change anything since we can simply set these charges to be trivial. If we were to consider the case $f\neq 1$, we would need to replace the $\Lambda^{[i]}$-tensor at the interface between the two subsystems by the corresponding 3-leg tensor, as indicated in Fig.~\ref{fig:MPS}\textbf{(b)}. However, since the main focus of this work is on the case $f=1$, we also restrict ourselves to this case here.

Let us now briefly mention a few important details regarding the anyonic generalization. Consider the Schmidt demcomposition of a state $|\psi \rangle$ at bond $i$,
\begin{align}
	|\psi\rangle = \sum_{c,j_c}\Lambda_{c,j_c}^{[i]} |\phi^A_{c,j_c}\rangle|\phi^B_{c,j_c}\rangle,
\end{align}
where $c$ denotes the anyonic charge at the interface between subsystems $A$ and $B$, $j_c$ the charge degeneracy index, $\Lambda_{c,j_c}^{[i]}$ the Schmidt values and $|\phi^{A/B}_{c,j_c}\rangle$ the Schmidt states. Then, the normalization constraint of $|\psi\rangle$ is given by~\cite{AnyonMPS}
\begin{align}
	\langle\psi | \psi \rangle = \sum_{c,j_c}d_c\left(\Lambda_{c,j_c}^{[i]}\right)^2=1,
\end{align}
where $d_c$ again denotes the quantum dimension of the anyonic charge $c$. The entanglement entropy corresponding to the bipartition into subsystems $A$ and $B$ can be expressed as~\cite{AnyonMPS}
\begin{align}
	S =  -\sum_{c,j_c}d_c\left(\Lambda_{c,j_c}^{[i]}\right)^2\ln\left(\left(\Lambda_{c,j_c}^{[i]}\right)^2\right).
\end{align}
At this point, we can readily see that a MPS with bond dimension $\chi=1$ can already correspond to an entangled state since a single $\Lambda_c^{[i]}=d_c^{-1/2}$ ensures correct normalization with entropy $S=\ln(d_c)$, which is nontrivial for non-abelian anyons $c$. This also confirms that a $\chi=2$ MPS can represent two anyon dimers, as claimed in Sec.~\ref{sec:ChargeResolvedALN} for the anyonic generalization of the AKLT state. When generalizing algorithms such as time-evolving block decimation (TEBD)~\cite{TEBD} and density matrix renormalization group (DMRG)~\cite{DMRG1, DMRG2} to anyonic MPS, in addition to keeping track of factors of quantum dimensions, one needs to introduce (3-leg) fusion tensors representing the anyonic fusion rules. These tensors need to be applied whenever tensors are to be reshaped as done, e.g., in order to perform a singular value decomposition. For more details on anyonic MPS in general, TEBD and DMRG for anyonic MPS, see Refs.~\cite{AnyonMPS, AnyonMPS2, AnyonMPS3, AnyonMPS4}.

Let us now focus on the computation of the ALN using MPS. Applying the definition of the APT in \eqref{eq:APT_def} to the MPS in Fig.~\ref{fig:MPS}\textbf{(a)} yields the operator shown in Fig.~\ref{fig:MPS}\textbf{(c)}. Here, we did not reverse the direction of the arrows associated with the anyons of subsystem $A$ (unlike in \eqref{eq:APT_def}). The hermitian conjugate $\left(\widetilde{\rho}^{T_A^a}\right)^\dagger$ can be obtained by taking the definition in \eqref{eq:APT_def}, reversing the braid direction and complex conjugating the prefactors. Since MPS in canonical form exploit the Schmidt decomposition, complex conjuagtion is trivial and the operator $\left(\widetilde{\rho}^{T_A^a}\right)^\dagger\widetilde{\rho}^{T_A^a}$ can be expressed as in Fig.~\ref{fig:MPS}\textbf{(d)}. Using the orthogonality properties of the canonical form, $\left(\widetilde{\rho}^{T_A^a}\right)^\dagger\widetilde{\rho}^{T_A^a}$ can be simplified to take the form in Fig.~\ref{fig:MPS}\textbf{(e)}, where we added the anyonic charge labels $c$ and $c'$ with respect to which $\left(\widetilde{\rho}^{T_A^a}\right)^\dagger\widetilde{\rho}^{T_A^a}$ is block diagonal (see \eqref{eq:BlockStructure2}). This can be used to compute both the ALN and the charge-resolved ALN. It can be seen that in Fig.~\ref{fig:MPS}\textbf{(e)}, only the $\Lambda^{[3]}$ tensors contribute to these quantities since the rest of the MPS corresponds to orthonormal states in the respective subsystems. This applies similarly to more general bipartitions, such that the ALN and charge-resolved ALN can be expressed as
\begin{align}
	&\mathcal{E}_a=\ln\left( \mathrm{Tr}\sqrt{\left(\widetilde{\rho}^{T_A^a}\right)^\dagger\widetilde{\rho}^{T_A^a}} \right)=\ln\left(\Bigg(\sum_{c,j_c}d_c\Lambda^{[i]}_{c,j_c}\Bigg)^2\right),\\
	&\mathcal{E}_a(c) = \ln\left( \Bigg( \sum_{j_c}d_c\Lambda^{[i]}_{c,j_c} \Bigg) \Bigg(\sum_{c',j_{c'}}d_{c'}\Lambda^{[i]}_{c',j_{c'}}\Bigg)  \right),
\end{align}
where $\Lambda^{[i]}$ denotes the tensor between subsystems $A$ and $B$ and $\Lambda^{[i]}_{c,j_c}$ the corresponding Schmidt values. Note that we use the charge labels $c$ and $c'$ instead of their anticharges. We do this to make the connection to the description of density matrices used in Sec.~\ref{sec:BlockStructure} and \eqref{eq:APT_sol}, which requires us to reverse the arrows of the fusion products in Fig.~\ref{fig:MPS}\textbf{(e)}, leading to $\overline{c}\rightarrow c$ and $\overbar{c'}\rightarrow c'$.

To compute the imbalance-resolved ALN, we need the third charge $g$ appearing in the block structure in \eqref{eq:BlockStructure2}, which is the fusion product of $c$ and $c'$. The connection can be made explicit via
\begin{align}
\vcenter{
\begin{tikzpicture}[line width=0.75pt]
\draw (0,0) -- (0,1.25) node[sloped,pos=0.5,allow upside down]{\arrowIn}; ; ;	
\node at (0,0.625) [anchor=east] {$c'$};
\draw (0.75,0) -- (0.75,1.25) node[sloped,pos=0.5,allow upside down]{\arrowIn}; ; ;	
\node at (0.75,0.625) [anchor=east] {$c$};
\end{tikzpicture}
}
\hspace{-195pt}
=\sum_g \sqrt{\frac{d_g}{d_{c'}d_c}}\hspace{-8pt}
\vcenter{
\begin{tikzpicture}[line width=0.75pt]
%\node at (0,1.25) [anchor=south] {$c'$};
%\node at (0,0) [anchor=north] {$c'$};
%\node at (0.75,1.25) [anchor=south] {$c$};
%\node at (0.75,0) [anchor=north] {$c$};
\node at (0+0.225,1.25-0.25) [anchor=east] {$c'$};
\node at (0+0.225,0+0.25) [anchor=east] {$c'$};
\node at (0.75-0.19,1.25-0.25) [anchor=west] {$c$};
\node at (0.75-0.19,0+0.25) [anchor=west] {$c$};
\node at (0.375,0.375+0.25) [anchor=east] {$g$};
\draw (0,0) -- (0.375,0.375) node[sloped,pos=0.5,allow upside down]{\arrowIn}; ; ;
\draw (0.75,0) -- (0.375,0.375) node[sloped,pos=0.5,allow upside down]{\arrowIn}; ; ;
\draw (0.375,0.375) -- (0.375,0.375+0.5) node[sloped,pos=0.5,allow upside down]{\arrowIn}; ; ;
\draw (0.375,0.375+0.5) -- (0,1.25) node[sloped,pos=0.5,allow upside down]{\arrowIn}; ; ;
\draw (0.375,0.375+0.5) -- (0.75,1.25) node[sloped,pos=0.5,allow upside down]{\arrowIn}; ; ;
\end{tikzpicture}
}\hspace{-200pt}.
\end{align}
Then, the imbalance-resolved ALN can be written as
\begin{align}
	\mathcal{E}_a^{\mathrm{imbalance}}(g) = \ln\left( \sum_{c,c'}N^g_{cc'}d_g \Bigg(\sum_{j_c,j_{c'}}\Lambda^{[i]}_{c,j_{c}}\Lambda^{[i]}_{c',j_{c'}}\Bigg)\right).
\end{align}

Note that the ALN and its charge- and imbalance-resolved versions can be evaluated with computational complexity $\mathcal{O}(\chi)$, where $\chi$ is the bond dimension of the MPS. This is however only valid for the bipartition geometry with trivial total fusion product $f=1$. In fact, we expect a much more unfavorable scaling when going beyond bipartitions, as seen for the BPT whose computational complexity is $\mathcal{O}(\chi^6)$ for \emph{tripartitions}~\cite{MPSPT}.

\bibliography{./references.bib}

\end{document}